\documentclass[a4paper,11pt]{article}
\pdfoutput=1
\usepackage{jcappub} 

\usepackage{slashed}

\allowdisplaybreaks
\newcommand{\fnl}{f_{\rm NL}}
\newcommand{\fnll}{f_{\mathrm{NL}}^{\mathrm{loc}}}
\newcommand{\fnle}{f_{\mathrm{NL}}^{\mathrm{equil}}}
\newcommand{\fnlo}{f_{\mathrm{NL}}^{\mathrm{orth}}}

\title{\boldmath Non-Gaussianity in  DHOST inflation}

\author[a]{Philippe Brax,}
\author[b]{Andrei Lazanu}

\affiliation[a]{Institut de Physique Th\'eorique, Universit\'e  Paris-Saclay, CEA, CNRS, F-91191 Gif-sur-Yvette Cedex, France}
\affiliation[b]{Laboratoire de Physique de l’Ecole normale sup\'erieure, ENS, Universit\'e PSL, CNRS, Sorbonne Universit\'e, Universit\'e de Paris, F-75005 Paris, France}

\emailAdd{philippe.brax@ipht.fr}
\emailAdd{andrei.lazanu@ens.fr}

\abstract{DHOST inflation models where deviations from a pure de Sitter background are induced by an axion-like potential can lead to large non-Gaussianities. We investigate the nature of non-Gaussianities in these models and compare to the results given by the \textit{Planck} experiment.  The overlap between the DHOST non-Gaussianities and the equilateral, orthogonal and local templates can be rendered arbitrarily small. On the other  hand, this does not preclude DHOST models from showing large non-Gaussianities as exemplified by their reduced bispectrum. As a result, they could be probed by future experiments and also  by  a more thorough analysis of the existing \textit{Planck} data. 
}

\begin{document}
\maketitle
\flushbottom

\section{Introduction}

The most general scalar-tensor theories involving one scalar degree of freedom are the Degenerate Higher Order Scalar Tensor (DHOST) theories \cite{Langlois:2015cwa, Achour:2016rkg, BenAchour:2016fzp, Crisostomi:2016czh, Crisostomi:2018bsp, Bombacigno:2021bpk}, generalising the Horndeski \cite{Horndeski1974} and beyond-Horndeski theories \cite{Zumalacarregui:2013pma,Gleyzes:2014dya, Gleyzes:2014qga}. These theories have been applied to  explain the accelerated expansion of the Universe and they avoid the appearance of Ostrogradski ghosts \cite{Ostrogradsky:1850fid} by adding degeneracy conditions \cite{Langlois:2015cwa, Motohashi:2016ftl, Motohashi:2017eya, Motohashi:2018pxg}. 

They can also be employed in the early Universe to build inflationary models. Inflation provides a source of primordial quantum fluctuations, which in turn generate matter perturbations and then become  structures \cite{Mukhanov:1981xt, Hawking:1982cz, Guth:1982ec, Starobinsky:1982ee}. The parameters connecting  inflation to  experiments such as the spectral index $n_s$ of scalar fluctuations have been measured with exquisite accuracy by Cosmic Microwave Background (CMB) probes such as the \textit{Planck} \cite{Akrami:2018odb} experiment, showing in particular a small departure from scale invariance for adiabatic scalar perturbations. This can be quantified at the level of the power spectrum through 
\begin{equation}
\mathcal{P}_{\zeta} (k) =  \mathcal{P}_{\zeta} (k_*)  \left(\frac{k}{k_*}\right)^{n_s-1} \, ,
\end{equation}
where $k_*$ is a pivot scale and $n_s$ is the spectral index.

DHOST theories lead to scale-invariant spectra, but extensions to such theories, including perturbations by  axion-like potential terms for instance yield small departures from scale-invariance in the scalar power spectrum, and at the same time are compatible with all the other inflationary constraints \cite{Brax:2021qlx}. This potential breaks the shift symmetry in field space and could result from non-perturbative dynamics. Contrary to traditional inflationary models with polynomial interactions, here the mass and quartic interactions are small perturbations to the background and do not drive inflation per se. They are only responsible for the breaking of scale invariance in the power spectra. We also emphasise  that these polynomial interactions do not break the degeneracy condition and therefore do not reintroduce a ghost in the spectrum of the theory. 

In \cite{Brax:2021qlx}, we have shown how to build such models and in particular we have investigated two scenarios: one corresponding to the current inflationary bounds, and another one corresponding to the tensor-to-scalar ratio that will come within the reach of future experiments, such as LiteBIRD \cite{Hazumi:2019lys}. We have, however,  used constraints arising from the two-point correlation functions (power spectra). In this work, we are extending the analysis of the axionic models described above, by looking at their scalar three-point correlation functions (bispectra), and we determine  their amplitudes by comparing them with current and future bounds for the non-Gaussian parameter $\fnl$ calculated for the local, orthogonal and equilateral templates of non-Gaussian fluctuations. In particular we find that these constraints can be easily satisfied although the primordial non-Gaussianiaties can be intrinsically large. This could for instance be detectable by future experiments and perhaps more significantly this could be constrained by a re-analysis of the Planck data. We intend to do this in the near future.

In section \ref{sec:bisp} we describe the formalism for calculating the bispectrum for DHOST theories starting from the Lagrangian, in section \ref{sec:constraints} we are providing the experimental constraints on different bispectrum templates, as well as the results we obtain for our models and we conclude in section \ref{sec:conclusions}. In two appendices, we give the expansion coefficients of the cubic Hamiltonian and the mode equation.

\section{The bispectrum of  DHOST models}
\label{sec:bisp}
\subsection{The models}
In this section, we  describe how to determine the bispectrum of primordial curvature perturbations in  DHOST theories, following our paper \cite{Brax:2021qlx} and using the methods developed in \cite{Gorji:2020bfl}. In the next few paragraphs, we will briefly describe the main features of the DHOST models perturbed by an axionic potential. As before, we restrict ourselves to the study of quadratic DHOST theories. Our aim is to calculate the bispectrum for these models.

The most general action involving up to second-order interactions in the scalar field can be written as
\begin{eqnarray}\label{action-DHOST}
S = \int d^4 x \sqrt{-g} \Big[ F_0(\phi,X) + F_1(\phi,X) \Box \phi + F_2(\phi,X) R 
+ \sum_{i=1}^5 A_i(\phi,X) L_i \Big] \,,
\end{eqnarray}
where $X=g^{\nu\eta}\phi_{\nu}\phi_{\eta}$, with $\phi_{\nu}\equiv\nabla_{\nu}\phi$,  the sign convention is $(-,+,+,+)$ and $L_i$ are all the five possible Lagrangians quadratic in the field $\phi$ and $A_i(\phi,X)$ their corresponding amplitudes with
\begin{align}\label{DHOST-L2s}
L_1 &= \phi_{\nu\eta} \phi^{\nu\eta} , \hspace{1cm} L_2 = (\Box \phi)^2 ,
\hspace{1cm} L_3 = \Box\phi \, \phi_{\nu}\phi^{\nu\eta} \phi_{\eta} , 
\nonumber \\
L_4 &= \phi^{\nu} \phi_{\nu\eta} \phi^{\eta\lambda}\phi_{\lambda} , 
\hspace{1cm} L_5 = (\phi_{\nu}\phi^{\nu\eta}\phi_{\eta})^2 \,.
\end{align}
We impose that $F_0>0$ and that it drives the expansion of the Universe in the inflationary background.
In order to be ghost-free and to satisfy the gravitational waves constraints, the functions $F_i$ and $A_i$ have to satisfy a set of degeneracy conditions \cite{Crisostomi:2017pjs, Crisostomi:2019yfo}.
We simplify the model assuming that the functions $F_i$ and $A_i$ only depend on the kinetic term $X$. Imposing the degeneracy conditions, the DHOST action becomes
\begin{eqnarray}\label{action-dhost}
S_{\rm D} = \int d^4 x \sqrt{-g} \bigg[ F_0(X) + F_1(X) \Box\phi + F_2(X) R
+ \frac{6 F_{2,X}^2}{F_2} \phi^{\nu} \phi_{\nu\eta} \phi^{\eta\lambda}\phi_{\lambda} \bigg]\,,
\end{eqnarray}
We will also consider perturbations by a potential interaction term, and we will  assume that their coefficients are small enough to be treated as  perturbations to the background cosmology driven by the DHOST action. 
Typically we will consider
\begin{equation}
    S_{\rm V} = -\int d^4 x \sqrt{-g}V(\phi)
\end{equation}
where $V(\phi)$ is an interaction, for instance of the form $V(\phi)= \mu^4 (\cos\frac{\phi}{f}-1)$ whose origin could be a non-perturbative breaking of the shift symmetry $\phi\to \phi+c$ like in the case of axions \cite{Marsh:2015xka}. By performing an expansion about $\phi=0$, this becomes
\begin{equation}
S_{\rm V} = \int d^4 \tilde x \sqrt{-\tilde g} \bigg[ - \frac{m_{}^2}{2}  \varphi^2 - \frac{\lambda_{}}{4!}  \varphi^4 \bigg] \,   
\end{equation}
at leading order. Notice that here $m^2<0$. The breaking of scale invariance is due to the non-vanishing $m^2$ and $\lambda$. 
We will write the action in terms of dimensionless coordinates and variables defined by:
\begin{equation}\label{coordinets-ch}
{\tilde t} \equiv \Lambda t \,, \hspace{1cm} {\tilde x}^i \equiv \Lambda x^i \,,
\end{equation}
\begin{equation}\label{dimensionless-couplings}
\phi \equiv M \, \varphi\,, \hspace{.5cm} X\equiv{M^2 \Lambda^2}{\mathrm x}\,, \hspace{.5cm} 
F_0 \equiv \Lambda^4 f_0 \,, \hspace{.5cm}
F_1 \equiv \frac{\Lambda^2}{M} f_1 \,, \hspace{.5cm} F_2 \equiv \Lambda^2 f_2 \,.
\end{equation}
where  we consider the models as low energy effective theories well below the Planck scale where quantum gravity effects should be considered. In the following we will choose $\Lambda\simeq m_{\rm Pl}$. Hence time and space are measured in Planck units. The scale $M$ gives the typical excursion scale of the scalar field and to avoid large excursion in units of the Planck scale, we will require that $M\ll m_{\rm Pl}$. The Hubble parameter is expressed in reduced units as $H = \Lambda h$.

At the  power spectrum level for primordial fluctuations, the models are characterised by the $\alpha$ parameters \cite{Bellini:2014fua, Gleyzes:2014rba, Langlois:2017mxy, Motohashi:2017gqb}
\begin{equation}\label{alpha-i}
\alpha_H \equiv - {\rm x} \frac{f_{2,{\rm x}}}{f_2}\,, \hspace{1cm}
\alpha_B \equiv \frac{1}{2} \frac{\dot{\varphi}\,{\rm x}}{h_b} \frac{f_{1,{\rm x}}}{f_2} + \alpha_H \,, \hspace{1cm}
\alpha_K \equiv - \frac{{\rm x}}{6h_b^2}\frac{f_{0,{\rm x}}}{f_2} + \alpha_H + \alpha_B \,,
\end{equation}
which are first order in derivative of the functions $f_i$, and the $\beta$ coefficients
\begin{eqnarray}\label{beta-K}
&&\beta_K \equiv - \frac{{\rm x}^2}{3} \frac{f_{0,{\rm x}{\rm x}} }{h_b^2 f_2} 
+ (1-\alpha_H) (1+3 \alpha_B) + \beta_B 
+ \frac{(1 + 6 \alpha_H - 3 \alpha_H^2) \alpha_K
- 2 ( 2 - 6 \alpha_H + 3 \alpha_K ) \beta_H}{1-3 \alpha_H} \,,
\nonumber \\
&& \beta_B \equiv  \dot{\varphi}\,{\rm x}^2 \frac{f_{1,{\rm x}{\rm x}}}{h_b f_2} \,,
\hspace{1cm}
\beta_H \equiv {\rm x}^2 \frac{f_{2,{\rm x}{\rm x}}}{ f_2}
\,,
\end{eqnarray}
which are second order in derivative of the functions $f_i$. We add the subscript \textit{dS} to the reduced Hubble parameter to emphasise that we are investigating de Sitter inflationary solutions. For the inflationary behaviour that we consider, the background solution of the equations of motion for the  scalar field is such that $\varphi=c-t$ and $x=-1$. More details can be found in \cite{Brax:2021qlx}.

\subsection{Non-Gaussianities}

In order to determine the bispectrum generated by such models, we follow the general prescription described in \cite{Maldacena2002} and extensively used in \cite{Chen:2010xka}. It consists in expanding the action to third order in the comoving gauge, to use the third order expansion to determine the Hamiltonian in the interation picture and then to calculate the three-point correlation function. 

As before, we work in the comoving gauge, where the line element for the scalar perturbations is then given by
\begin{equation}\label{metric-FRW-Perturbations}
ds^2 = \Lambda^2 \Big( - ( 1 + 2 A ) d{\tilde t}^2 
+ 2 {\tilde \partial}_i B d{\tilde t} d{\tilde x}^i 
+ a^2 ( 1 + 2 \psi ) \delta_{ij} d{\tilde x}^i d{\tilde x}^j \Big) \,,
\end{equation}
where $(A,B,\psi)$ are scalar perturbations depending on the dimensionless coordinates $({\tilde t},{\tilde x}^i)$ and ${\tilde \partial}_i$ is the derivative with respect to ${\tilde x}^i$. 

Substituting (\ref{metric-FRW-Perturbations}) in (\ref{action-DHOST}), expanding the action up to the cubic order in the perturbations and integrating by parts in time, the third order action for the DHOST theory is obtained 
\begin{equation}\label{action-SS-bare}
S^{(3)}_{\rm D} \equiv \int d{\tilde t} d^3{\tilde x} 
 {\tilde {\cal L}}_{\rm D}^{(3)}(\psi,\dot{\psi},A, \dot{A},B,\dot{B} ) \, ,
\end{equation}
where only the time dependence was explicitly expressed. Gradients up to second order of each of the perturbation variables appear explicitly. As a difference to our previous work~ \cite{Brax:2021qlx}, here we do not work in Fourier space. We express the action in terms of the comoving curvature perturbation $\zeta$  \cite{Riotto2002},
\begin{equation}\label{xi-def}
\zeta \equiv \psi + \alpha_H A \,.
\end{equation}
The two fields
$A$ and $B$ which can be treated as Lagrange multipliers in the second order action \cite{Brax:2021qlx} have the following expressions 
\begin{equation}\label{Phi-B-sol}
A = \frac{1}{1+\alpha_B} \frac{\dot{\zeta}}{h_b} \,, \hspace{1cm}
B =  3 \bigg[ 1 - \frac{\beta_K}{(1+\alpha_B)^2} \bigg] a^2 \,  \nabla^{-2} \dot{\zeta}
- \frac{1+\alpha_H}{1+\alpha_B} \, \frac{\zeta}{h_b} \,,
\end{equation}
where the inverse Laplacian is defined as $\nabla^{-2} (\nabla^2 P) = P$. This action now only depends on the $\zeta$ and $\dot{\zeta}$. In order to proceed, we need to express it in terms of fields at different space positions and convert it to Fourier space, replacing $\partial_\alpha \to i k_{\alpha}$. Prior to this replacement, the number of types of operators is of order 80. To illustrate this procedure, we take the example of the operator below, which is transformed in Fourier space into 
\begin{equation}
\partial_\alpha \partial_{\beta} (\nabla^{-2} \zeta') \partial^{\alpha}(\nabla^{-2} \zeta')\partial^{\beta} \zeta \to \frac{(\mathbf{k}_1 \cdot \mathbf{k}_2)(\mathbf{k}_1 \cdot \mathbf{k}_3)}{k_1^2 k_2^2)} \zeta_1' \zeta_2' \zeta_3 + {5 \text{\, perm.}} \,
\end{equation}
followed by the use of the cosine theorem to replace the scalar product of the two wavevectors with the magnitudes of $k_1$, $k_2$ and $k_3$.
All the terms arising from the third order expansion of the perturbation $S_V$ are of the same type.
Hence, the cubic action can be written as
\begin{align}
   S_3= \int d\eta &(\prod_{i=1}^3 \slashed{d}^3\tilde k_i) \slashed{\delta} (\vec {\tilde k}_1 + \vec {\tilde k}_2 + \vec {\tilde k}_3) (C_0\zeta (\tilde k_1) \zeta (\tilde k_2) \zeta (\tilde k_3) + C_1\zeta^\prime (\tilde k_1) \zeta (\tilde k_2) \zeta (\tilde k_3) \nonumber \\ 
   &+C_2\zeta^\prime (\tilde k_1) \zeta^\prime (\tilde k_2) \zeta (\tilde k_3)
   +C_3\zeta^\prime (\tilde k_1) \zeta^\prime (\tilde k_2) \zeta^\prime (\tilde k_3))
\end{align}
where the four coefficients have been determined explicitly and depend on $\eta$ and the magnitudes of the three wavevectors. They are given in Appendix \ref{app1}. These are the only operators that appear at this order. 
Typically we are interested in the three point functions
\begin{equation}
    \langle 0\vert \zeta (\tilde k_1) \zeta(\tilde k_2) \zeta (\tilde k_3)\vert 0\rangle = - i\int d\eta\langle 0\vert  [\zeta (\tilde k_1) \zeta(\tilde k_2) \zeta (\tilde k_3), H_3 ]\vert 0\rangle \, ,
\end{equation}
where the interaction picture Hamitonian is given by \cite{Weinberg:2005vy}
\begin{align}
H_3=-\int (\prod_{i=1}^3 \slashed{d}^3\tilde k_i) & \slashed{\delta} (\vec {\tilde k}_1 + \vec {\tilde k}_2+ \vec {\tilde k}_3) (C_0\zeta (\tilde k_1) \zeta (\tilde k_2) \zeta (\tilde k_3)  + C_1\zeta^\prime (\tilde k_1) \zeta (\tilde k_2) \zeta (\tilde k_3) \nonumber \\
&+ C_2\zeta^\prime (\tilde k_1) \zeta^\prime (\tilde k_2) \zeta (\tilde k_3)
+ C_3\zeta^\prime (\tilde k_1) \zeta^\prime (\tilde k_2) \zeta^\prime (\tilde k_3)) \,.
\label{cub}
\end{align}
Using the result \cite{Chen:2010xka} obtained from the time evolution of operators in the interaction picture
\begin{equation}
\langle0\vert  W(t) \vert 0\rangle = {\rm Re}\left[\langle0\vert  -2i W^I(t) \int_{-\infty (1+i \epsilon)}^t H^{I}(t') dt' \vert 0\rangle\right] \,,
\end{equation}
together with Wick's theorem, and the mode quantisation described in \cite{Brax:2021qlx}, the three point function decomposes into a sum of four terms coming from the four terms in the interaction Hamiltonian at cubic order (\ref{cub}), i.e.
\begin{equation} 
\langle 0\vert \zeta (\tilde k_1) \zeta(\tilde k_2) \zeta (\tilde k_3)\vert 0\rangle =\sum_{i=0}^3 B_0(k_1,k_2,k_3,\eta_f)
\end{equation}
evaluated at $\eta_f$. The choice of $\eta_f$ will be discussed below.  The 
expressions for the four terms become,
\begin{align}
B_0&(k_1,k_2,k_3,\eta_f)=-\mathrm{Re}\left[-2i \zeta(\mathbf{k}_1,\eta_f)\zeta(\mathbf{k}_2,\eta_f)\zeta(\mathbf{k}_3,\eta_f) \int \frac{d^3 q_1}{(2 \pi)^3} \frac{d^3 q_2}{(2 \pi)^3} \frac{d^3 q_3}{(2 \pi)^3}  \right. \nonumber \\
&\times \left.\int_{-\infty (1-i \epsilon)}^{\eta_f} d \eta a C_0 e^{-i(\mathbf{q}_1+\mathbf{q}_3+\mathbf{q}_2)\cdot \mathbf{x}} \mathbf{q}_1,\eta)\zeta(\mathbf{q}_2,\eta)\zeta(\mathbf{q}_3,\eta) \right] \nonumber \\
& = -\mathrm{Re}\left[-2i  \int_{-\infty (1-i \epsilon)}^{\eta_f} d \eta a C_0   u(k_1,\eta_f)u(k_2,\eta_f)u(k_3,\eta_f) u^*(k_1,\eta)u^*(k_2,\eta)u^*(k_3,\eta)  \right] + 5 \mathrm{\,perm.}
\label{eq:B0}
\end{align}

\begin{align}
B_1(k_1,k_2,k_3,\eta_f)&= -\mathrm{Re}\left[-2i  \int_{-\infty (1-i \epsilon)}^{\eta_f} d \eta a C_1   u(k_1,\eta_f)u(k_2,\eta_f)u(k_3,\eta_f) \right. \nonumber \\
&\qquad \qquad \times \left. u^{\prime*}(k_1,\eta)u^*(k_2,\eta)u^*(k_3,\eta)  \right] + 5 \mathrm{\,perm.} \\
B_2(k_1,k_2,k_3,\eta_f)&= -\mathrm{Re}\left[-2i  \int_{-\infty (1-i \epsilon)}^{\eta_f} d \eta a C_2   u(k_1,\eta_f)u(k_2,\eta_f)u(k_3,\eta_f) \right. \nonumber \\
&\qquad \qquad \times \left. u^{\prime*}(k_1,\eta)u^{\prime*}(k_2,\eta)u^*(k_3,\eta)  \right] + 5 \mathrm{\,perm.} \\
B_3(k_1,k_2,k_3,\eta_f)&= -\mathrm{Re}\left[-2i  \int_{-\infty (1-i \epsilon)}^{\eta_f} d \eta a C_3   u(k_1,\eta_f)u(k_2,\eta_f)u(k_3,\eta_f) \right.\nonumber \\
&\qquad \qquad \times \left. u^{\prime*}(k_1,\eta)u^{\prime*}(k_2,\eta)u^{\prime*}(k_3,\eta)  \right] + 5 \mathrm{\,perm.} 
\label{eq:B3}
\end{align}
The final bispectrum is obtained by summing up the four preceding expressions. 
The DHOST bispectrum depends on six additional parameters compared  to the ones determining the scalar power spectrum: $f_{2,{\rm xxx}}$, $f_{0,{\rm xxx}}$,  $f_{1, {\rm xxx}}$, $f_1$, $\beta_B$ and $\beta_H$, while the bispectrum arising from the axionic perturbation to the model is independent of these parameters. The six parameters can be freely chosen. Here the parameters $\beta_B$ and $\beta_H$, described in \cite{Gorji:2020bfl}, quantify the magnitudes of $f_{1,{\rm xx}}$ and $f_{2, {\rm xx}}$ respectively.
Finally and in order to capture all the relevant information, we choose $\eta_f=-\frac{1}{c_s \max(k_1,k_2,k_3)}$. {{This choice, analogous to that for the power spectrum in \cite{Brax:2021qlx}, is justified by the requirement that all the modes have  to be out the horizon and that the non-Gaussianities are evaluated when all the modes have crossed the horizon.}} The modes functions $u$ appearing in the expansion of $\zeta$ in terms of annihilation and creation operators are given by
\begin{align}
u({\tilde k},\eta)=\frac{v({\tilde k},\eta)}{z(\eta)}
\end{align}
where
\begin{align}
\label{DHOST:scalar}
v({\tilde k},\eta)&=\frac{1}{\sqrt{2 \bar{c}_s {\tilde k}}}\left(1-\frac{i}{\bar{c}_s {\tilde k}\eta} \right) \exp(-i \bar{c}_s {\tilde k}\eta) \,,\\
z^2(\eta)&=\frac{6 f_2}{h_{\mathrm{dS}}^2 \eta^2} \left(1-\frac{\beta_K}{(1+\alpha_B)^2} \right) \,,
\end{align}
for the DHOST case. When considering the perturbed model, $v$ and $z$ are modified, as shown in Eqs. (3.11)-(3.14) of \cite{Brax:2021qlx}. This is enough to evaluate the bispectrum. 

\section{Constraints on primordial non-Gaussianity}
\label{sec:constraints}
\subsection{Experimental limits on primordial non-Gaussianity}
Depending on the inflationary model considered, the primordial gravitational potential and the resulting non-Gaussianities can be produced in several shapes, the most popular being the local, equilateral and orthogonal templates, which have the following primordial bispectra
\begin{align}
B_{\Phi}^{\text{loc}}&(k_1,k_2,k_3)=2 \left[P_{\Phi}(k_1)P_{\Phi}(k_2)+\text{2 perms} \right] \, ,\\
B_{\Phi}^{\text{equil}}&(k_1,k_2,k_3)=6 \left\{-[P_{\Phi}(k_1)P_{\Phi}(k_2)+\text{2 perms} ]\right. \nonumber \\
&-2[P_{\Phi}(k_1)P_{\Phi}(k_2)P_{\Phi}(k_3)]^{2/3} 
+[P_{\Phi}^{1/3}(k_1)P_{\Phi}^{2/3}(k_2)P_{\Phi}(k_3)+\text{5 perms}]\left. \right\} \,, \\ 
  B_{\Phi}^{\text{orth}}&(k_1,k_2,k_3) = 6\big[ 3(P_\Phi^{1/3}(k_1)P_\Phi^{2/3}(k_2)P_\Phi(k_3)+5\text{ perms}) \nonumber  \\
  &-3 \left[P_{\Phi}(k_1)P_{\Phi}(k_2)+\text{2 perms} \right] 
  -8 (P_\Phi(k_1)P_\Phi(k_2)P_\Phi(k_3))^{2/3}
\big] \,.
\end{align} 
If the bispectrum of a given model, for instance the axionic ones considered here,  has one of the above primordial shapes, then the parameter $\fnl$ represents the amplitude of its non-Gaussianities.
\textit{Planck} has  placed the tightest constraints so far on the amplitudes of these shapes using the CMB (Cosmic Microwave Background), with the latest limits being 
\begin{align}
-11.1<f_{\rm NL}^{\rm local} &<9.3 \\
-120<f_{\rm NL}^{\rm equil} &< 68 \\
-86<f_{\rm NL}^{\rm orth} &<10 
\label{eq:limits}
\end{align}
at 95 \%CL \cite{Akrami:2019izv}. Large-scale structure probes have already been used to constrain these shapes \cite{Scoccimarro2001b,Feldman2001,Verde2002,Marin2013,GilMarin2014,GilMarin2017}, but these constraints are not competitive compared to those from the CMB. However, bispectrum results from the large-scale structures have the potential to improve over existing CMB bounds \cite{Karagiannis:2018jdt}, at least for the local shape.

The constraints presented above all assume that the bispectrum generated by the inflationary model has one of the three typical shapes. If that is not the case, one should in principle investigate the constraints on the shape defined by a particular model such as the axionic models considered in this paper  using a new analysis of the \textit{Planck} data. This is left for future work.  In order to avoid this complication in this paper, a simplified method has been proposed in \cite{Babich:2004gb}, based on the shape correlators. We follow the formalism described in Ref. \cite{Fergusson:2008ra} in order to calculate the scalar products corresponding to the projection of a given model on the three templates. First one defines the shape function of a given bispectrum  by 

\begin{equation}
S(k_1,k_2,k_3)=  k_1^2 k_2^2 k_3^2 B_{\Phi}(k_1,k_2,k_3)   \, .
\end{equation}
Then, on can introduce the scalar product between two bispectrum shapes to be
\begin{equation}
S_1 \cdot S_2 = \int_{\nu_k} S_1(k_1,k_2,k_3) S_2(k_1,k_2,k_3) \omega(k_1,k_2,k_3) d\nu_k 
\end{equation}
where $\nu_k$ is the three-dimensional region satisfying $k_1,k_2,k_3 \in [k_{\rm min},k_{\rm max}]$ and the triangle conditions $k_1+k_2 \ge k_3$, $k_2+k_3 \ge k_1$ and $k_3+k_1 \ge k_2$ and the weighting function $\omega$ is taken to be
\begin{equation}
\omega(k_1,k_2,k_3)=\frac{1}{k_1+k_2+k_3}   \, . 
\end{equation}
Following the arguments of Ref. \cite{Babich:2004gb}, we then define  the \textit{fudge factors} $F^i$,
\begin{equation}
\label{fudge}
F^i=\frac{S\cdot S^i}{S^i \cdot S^i}\, ,
\end{equation}
where $i$ stands for local, equilateral and orthogonal and $S$ is the shape of the bispectrum in the model of interest. The projection $F^i$ quantifies the correlation between the model and each of the three primordial shapes and allows one to restrict oneself to the equilateral configuration of the bispectrum for the given shape. This projection modifies the allowed bounds obtained from experiments, 
\begin{equation}
\label{eq:minmax}
\min<\fnl^i(k_1,k_2,k_3)<   \max \to \frac{\min}{F^i}<\fnl^i (k,k,k)<   \frac{\max}{F^i} \,.
\end{equation}
This technique allows one to focus only on equilateral configurations  for each of the primordial shapes and to tune the parameters of any  given  model such that the constraints from {\it Planck} are satisfied. 

\subsection{Pure DHOST models}
In the DHOST case with no axionic perturbations, we consider two models, as in \cite{Brax:2021qlx}: one compatible with the current constraints for the tensor-to-scalar ratio, $r=0.04$, characterised by the parameters: $\alpha_B=1$, $\alpha_H=1.04$ and $\beta_K=3.97343$,  $f_2=2.7$ and $h_{\mathrm{ds}}=3 \times 10^{-5}$, and a second one that corresponds to constraints predicted by future CMB experiments, such as LiteBIRD \cite{Hazumi:2019lys} of $r \sim 10^{-3}$ which has parameters $\alpha_B=1$, $\alpha_H=1.001$, $\beta_K=3.9993$, $h_{\mathrm{ds}}=10^{-5}$ and $f_2=8.8$. 

Using the prescriptions from the above section, in order to find viable inflationary models compatible with the \textit{Planck} results (for the first model) and for future experiments (for the second model), it is sufficient to find parameters that fix $\fnll$, $\fnle$ and $\fnlo$ close to 0, while ensuring that the \textit{fudge factors} (\ref{fudge}) are sufficiently close to 1.
Then, for the first model, the three $\fnl$ parameters are given by
\begin{align}
\fnll&=\frac{B(k,k,k)}{B_{\Phi}^{\rm loc}(k,k,k)} =-2845.68 + 26.964 f_{2,{\rm xxx}} \nonumber \\
&+ 2.496\times 10^9 f_{0,{\rm xxx}} + 3.889 \times 10^6 f_1 + 
 224701 f_{1,{\rm xxx}} - 103.838 \beta_B + 736.024 \beta_H \\
 \fnle&=\frac{B(k,k,k)}{B_{\Phi}^{\rm equil}(k,k,k)}=3 \fnll\\
 \fnlo&=\frac{B(k,k,k)}{B_{\Phi}^{\rm orth}(k,k,k)} = 3 \fnll
\end{align}
Notice that in this DHOST scenario these quantities are scale-invariant and that the equalities above are only valid in the equilateral configuration of the triangles. These results imply that there is only one constraint coming from $\fnl \approx 0$ and three from  fixing the fudge factors $F^i = \mathcal{O}(1)$. As a result  the parameter space for DHOST models is relatively large. Hence the bispectrum does not place  significant constraints on this class of theories.

As an illustrative example, for $f_{2,{\rm xxx}}=0$, $f_{0,{\rm xxx}}=0$,  $f_{1, {\rm xxx}}=0$, $f_1=0$, $\beta_B=0$ and $\beta_H=3.86629$, we get $\fnll=3.2 \times 10^{-13}$, $\fnle=\fnlo=9.5 \times 10^{-13}$ and $F^{\rm loc}=0.93$, $F^{\rm equil}=0.35$, $F^{\rm orth}=-0.16$, all compatible with the \textit{Planck} constraints.
For the second model with a smaller tensor-to-scalar ratio, we get
\begin{align}
 \fnle&=\fnlo =3 \fnll = 3 (-202362 + 615.355 f_{2,{\rm xxx}} + 5.1279 \times 10^{11} f_{0,{\rm xxx}} \nonumber \\
 &+ 2.6263 \times 10^8 f_1 + 
 1.53839 \times 10^7 f_{1,{\rm xxx}} - 7566.88 \beta_B + 54369.1 \beta_H) \, .
\end{align}
As the constraints on $\fnll$ are likely to be improved by a factor of at least 20 in the next decade \cite{Karagiannis:2018jdt}, we fix $\fnl \approx 0$ as for this second model.
For $f_{2,{\rm xxx}}=0$, $f_{0,{\rm xxx}}=0$,  $f_{1, {\rm xxx}}=0$, $f_1=0$, $\beta_B=0$ and $\beta_H=3.722$, we get $\fnll=2.0 \times 10^{-11}$, $\fnle=\fnlo=6.1 \times 10^{-11}$ and $F^{\rm loc}=-264$, $F^{\rm equil}=-1384$, $F^{\rm orth}=1665$, compatible with future $\fnl$ constraints.

\subsection{DHOST models with perturbations}

When we consider DHOST models perturbed by an axionic potential, 
we notice that the perturbations to the non-Gaussianities appear in two places, i.e.  they modify  the mode functions $u$ and also the interaction Hamiltonian.  Let us analyse these two possibilities.

\subsubsection{Mode functions}
Let us first consider the perturbed mode functions and then  estimate the magnitude of their contribution to the bispectrum.
We consider small perturbations of the mode functions $v$ and $z$ around the DHOST solution $v_0$ and $z_0$ respectively,
\begin{align}
v (k,\eta)=v_0(k,\eta) [1+\lambda v_{\lambda}(k,\eta)+m^2 v_{m^2}(k,\eta)] \, , \\
z (k,\eta)=z_0(k,\eta) [1+\lambda z_{\lambda}(k,\eta)+m^2 z_{m^2}(k,\eta)] \, ,
\end{align}
where $v$ satisfies the mode equation
\begin{equation}
v'' + k^2K^2 v =0 \, ,
\end{equation}
as in Eq. (3.11) of \cite{Brax:2021qlx}. Here the prime denotes derivatives with respect to $\eta$ and the function $K^2$ can be written as
\begin{equation}
K^2=K_0^2+\lambda K_{\lambda}^2(k,\eta)+m^2 K_{m^2}^2(k,\eta) \, 
\end{equation}
with
\begin{align}
K_0^2(x)=-\frac{(\alpha_B+1) (\alpha_B-\alpha_H)}{3 \left(\alpha_B^2+2 \alpha_B-\beta_K+1\right)}-\frac{2}{x^2} 
\end{align}
and $K_{\lambda}^2$ and $K_{m^2}^2$ are given in the Appendix \ref{app2}.
Then $v_{\lambda}$ satisfies
\begin{equation}
v_{\lambda}''+\frac{v_0'}{v_0}v_{\lambda}'+k^2 K_{\lambda}^2=0 \,
\end{equation}
which can be expressed as a set of two coupled first-order ODEs. The solution for the derivative of $v_{\lambda}$ is given by
\begin{equation}
v_{\lambda}'(\eta)=\frac{-\int^{\eta} \frac{k^2 K_\lambda^2 (\eta')}{v_0^2(\eta')} d\eta'}{v_0^2(\eta)}     
\label{eq:vp}
\end{equation}
and $v_{\lambda}$ can be obtained by integrating this result. A similar solution can be written for $v_{m^2}$.

Hence, the perturbed mode functions $u$ can be expressed at first order in the perturbation variables as
\begin{align}
u(k,\eta) &\equiv u_{\rm DHOST}(k,\eta) + \Delta u(k,\eta)\nonumber \\
&= u_{\rm DHOST}(k,\eta) [1+ \lambda (v_{\lambda}(k,\eta)-z_{\lambda}(k,\eta))+m^2  (v_{m^2}(k,\eta)-z_{m^2}(k,\eta)] \,.
\label{eq:uall}
\end{align}
When computing the correlation functions, we have products of six instances of $u$ (or $u^*$). Hence, when computing perturbations at first order, we only need to perturb one of the $u$'s at a time, keeping the others unperturbed and summing over the six contributions. 
The integrals of the type described in Eq. (\ref{eq:vp}) can no longer be performed analytically, and therefore the final integral  needs to be evaluated numerically. The integrals are oscillating and to avoid the uncertainties related to introducing a hard cutoff at early times, we multiply the integrand by a damping factor using the prescription described in Refs. \cite{Chen:2006xjb,Chen:2008wn,Zhang:2020uek}. The integrals corresponding to the corrections modify Eqs. (\ref{eq:B0})-(\ref{eq:B3}) to
\begin{align}
\Delta B_i&(k_1,k_2,k_3,\eta_f)= -\mathrm{Re}\left[-2i  \int_{-\infty (1-i \epsilon)}^{\eta_f} d \eta a C_0   \Delta u(k_1,\eta_f)u(k_2,\eta_f)u(k_3,\eta_f) \right. \nonumber \\
& \left. \times u^{a*}(k_1,\eta)u^{b*}(k_2,\eta)u^{c*}(k_3,\eta)  \right] + \mathrm{\,perms.} \, ,
\label{eq:deltaB0bis}
\end{align}
where $a, b, c \in \{\varnothing,'\}$ based on whether $u$ or $u'$ is required.

\begin{figure}[!h]
\centering
\includegraphics[width=0.55\linewidth]{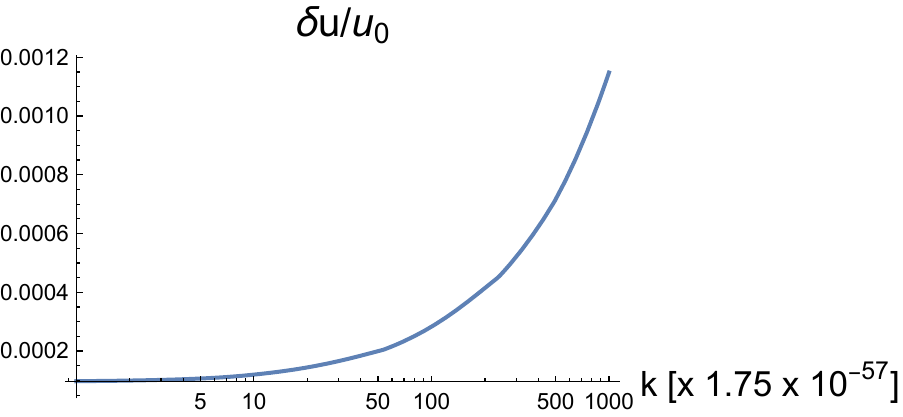}
\caption{Corrections to the modes $u$ from the perturbations. We can see that they are negligible at the precision level we are working. The parameters are chosen to match those from the first model: $\alpha_B=1$, $\alpha_H=1.04$ and $\beta_K=3.97343$,  $f_2=2.7$ and $h_{\mathrm{ds}}=3 \times 10^{-5}$. }
\label{fig:modespert}
\end{figure}

After performing the numerical integrations, we  observe that the value of the corrections from $u$ give a negligible contribution to the mode functions  and then to the bispectrum, of the order of $10^{-4}-10^{-3}$ compared to the DHOST bispectrum. This is illustrated in Fig. \ref{fig:modespert}, where we have plotted the contribution of the perturbation of the mode function $u$ ($u(k,-1/k)- u_{\rm DHOST}(k,-1/k))/u_{\rm DHOST}(k,-1/k))$ showing that this is indeed several orders of magnitudes smaller then 1.
As we will explain in the next section, this is significantly  smaller then the corrections coming from the Hamiltonian.

\subsubsection{Hamiltonian}
\begin{figure}[!h]
\centering
\includegraphics[width=0.49\linewidth]{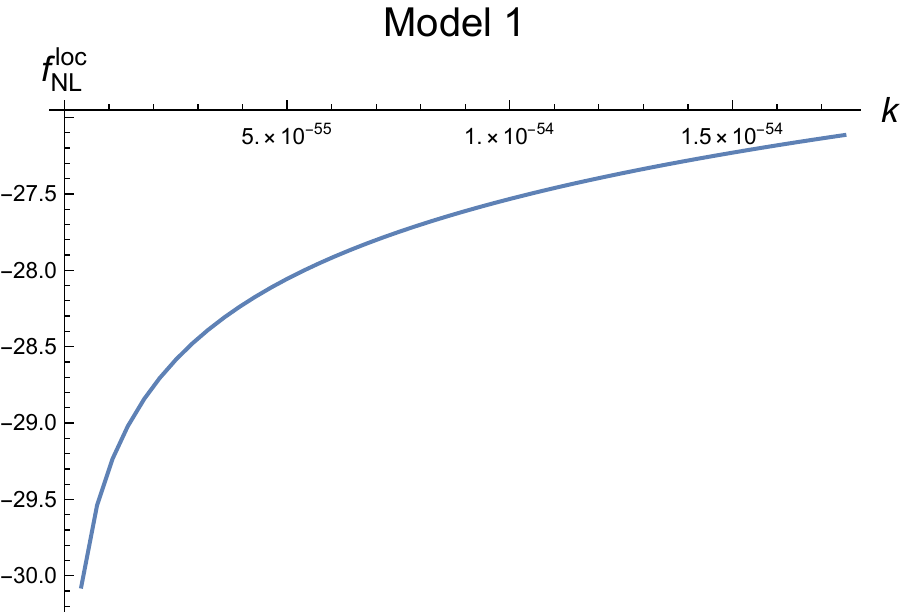}
\includegraphics[width=0.49\linewidth]{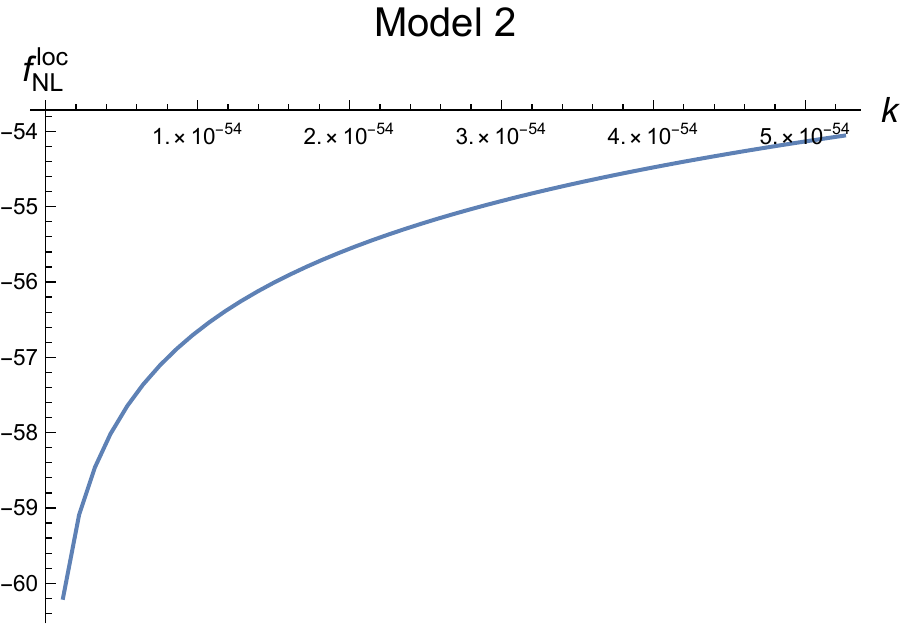} 
\caption{The scale dependent $\fnll$ coming from the perturbations to the Hamiltonian for the two models considered (left and right) . We have  $\fnlo=\fnle= 3 \fnll$ in the equilateral configuration { as a function of $k$ in Planck scale  units}. { These non-Gaussianities are compatible with the {\it Planck} constraints.} The first model ($r=0.04$) has parameters: $\alpha_B=1$, $\alpha_H=1.04$ and $\beta_K=3.97343$,  $f_2=2.7$, $h_{\mathrm{ds}}=3 \times 10^{-5}$, $f_{0,{\rm xxx}}= 2 \times 10^{-6}$,  $f_{1, {\rm xxx}}=-0.16$, $f_{2,{\rm xxx}}=150$, $f_1=0.0076$, $\beta_B=0$ and $\beta_H=0$; the second model ($r=10^{-3}$) has parameters: $\alpha_B=1$, $\alpha_H=1.001$, $\beta_K=3.9993$, $h_{\mathrm{ds}}=10^{-5}$, $f_2=8.8$, $f_{0,{\rm xxx}}= 7 \times 10^{-6}$,  $f_{1, {\rm xxx}}=0.12$, $f_{2,{\rm xxx}}=-2675$, $f_1=-0.013$, $\beta_B=0$ and $\beta_H=0.1$.}
\label{fig:bng}
\end{figure}

The perturbations to the pure DHOST case also introduce additional terms in the interaction Hamiltonian.
In this case, as these are already at the perturbation level, we do not need to perform any perturbation on the mode functions $u$, and we can therefore calculate the analytical contributions of the perturbations using the DHOST mode functions when calculating non-Gaussianities. 
The coefficients of the terms appearing in the interaction Lagrangian and then in the interaction Hamiltonian are
\begin{align}
C_0^{\rm pert}&=\frac{a (c-t)^2 \left(c^2 \lambda-2 c \lambda t+\lambda t^2+12 m^2\right) \left(6 a^2 (\alpha_B+1)^2 H^2-(\alpha_H+1)^2 \left(k_1^2+k_2^2+k_3^2\right)\right)}{48 (\alpha_B+1)^2 H^2} \\
C_1^{\rm pert}&=\frac{a (\alpha_H+1) (c-t)^2 \left(c^2 \lambda-2 c \lambda t+\lambda t^2+12 m^2\right)}{48 (\alpha_B+1)^3 H^3}
 \nonumber \\
&\qquad \times \left((\alpha_H+1)^2 \left(k_1^2+k_2^2-k_3^2\right)-6 a^2 H^2 \left(2 \alpha_B^2+4 \alpha_B-\beta_K+2\right)\right)\\
C_2^{\rm pert}&=\frac{a^3 (c-t)^2 \left(c^2 \lambda-2 c \lambda t+\lambda t^2+12 m^2\right)}{16 (\alpha_B+1)^4 H^2 k_2^2 k_3^2} \left[ 3 a^2 H^2 \left(\alpha_B^2+2 \alpha_B-\beta_K+1\right)^2 \left(k_1^2-k_2^2-k_3^2\right) \right.\nonumber \\
&\qquad +(\alpha_H+1)^2 \left(k_1^2 \left(\alpha_B^2+2 \alpha_B-\beta_K+1\right) \left(k_2^2+k_3^2\right)-k_2^4 \left(\alpha_B^2+2 \alpha_B-\beta_K+1\right) \right. \nonumber \\
&\qquad \left.\left.+2 k_2^2 k_3^2 \left(2 \alpha_B^2+4 \alpha_B-\beta_K+2\right)-k_3^4 \left(\alpha_B^2+2 \alpha_B-\beta_K+1\right)\right) \right]\\
C_3^{\rm pert}&=-\frac{a^3 (\alpha_H+1) (c-t)^2 \left(c^2 \lambda-2 c \lambda t+\lambda t^2+12 m^2\right) }{16 (\alpha_B+1)^5 H^3 k_1^2 k_2^2 k_3^2}  \left[3 a^2 H^2 \left(\alpha_B^2+2 \alpha_B-\beta_K+1\right)^2 \right . \nonumber \\
& \qquad  \times \left. \left(-2 k_1^2 \left(k_2^2+k_3^2\right)+k_1^4+\left(k_2^2-k_3^2\right)^2\right)+2 (\alpha_B+1)^2 (\alpha_H+1)^2 k_1^2 k_2^2 k_3^2\right]
\end{align}
These coefficients are independent of the six additional DHOST parameters $f_{2,{\rm xxx}}$, $f_{0,{\rm xxx}}$,  $f_{1, {\rm xxx}}$, $f_1$, $\beta_B$ and $\beta_H$, and yield scale dependent bispectra that are fixed for each model. In Figure \ref{fig:bng} we represent the three $\fnl$ coefficients in the equilateral configurations of the triangles for the two models. 
We notice that the corrections to the pure DHOST case yield significant deviations from zero for both models.

For the first model, in order to find a total bispectrum that is compatible with the \textit{Planck} data, we choose parameters for the pure DHOST model such that its bispectrum has $f_{\mathrm{NL}\, (\rm DHOST)}^{\mathrm{loc}} =27.5$ and $f_{\mathrm{NL}\, (\rm DHOST)}^{\mathrm{equil}} = f_{\mathrm{NL}\, (\rm DHOST)}^{\mathrm{orth}}=54$. Then, the resulting $\fnl$ parameters are $\fnll \in [-2.5,0.5]$ and $\fnle, \fnlo \in [-36, -28]$ after choosing  $f_{0,{\rm xxx}}= 2 \times 10^{-6}$,  $f_{1, {\rm xxx}}=-0.16$, $f_{2,{\rm xxx}}=150$, $f_1=0.0076$, $\beta_B=0$ and $\beta_H=0$. These parameters give the overall fudge factors of $F^{\rm loc}=2$, $F^{\rm equil}=1$, $F^{\rm orth}=1$. We note that, while the parameters of the models have been fine tuned for these specific shapes to make the corresponding bispectra compatible with \textit{Planck} measurements, the overall bispectrum is large and is potentially detectable.
For the second model, corresponding to future observations, we assume that the future measurements of the three $\fnl$'s will be close to zero, and therefore we choose parameters such that $\fnll \in [-3,3]$, $\fnle, \fnlo \in [9,9]$. We require fudge factors of $F^{\rm loc}=0.03$, $F^{\rm equil}=0.1$ and $F^{\rm orth}=0.1$ in order to get models compatible with $\Delta \fnll \sim 0.1$ and $\Delta \fnle, \Delta \fnlo,  \sim 1$. Parameters yielding such as result are $f_{0,{\rm xxx}}= 7 \times 10^{-6}$,  $f_{1, {\rm xxx}}=0.12$, $f_{2,{\rm xxx}}=-2675$, $f_1=-0.013$, $\beta_B=0$ and $\beta_H=0.1$. We note that the parameters of the models can be tuned further to match even tighter measurements.
Moreover, this choice might not be unique as out of the six parameters, we need to fix only four conditions. Fixing the condition that $\fnl$ is small for the three templates yields a linear relation between the parameters, but the scalar products are non-linear functions of the parameters, and hence fixing the fudge factors requires considerable fine-tuning. 

\begin{figure}[!h]
\centering
\includegraphics[width=0.49\linewidth]{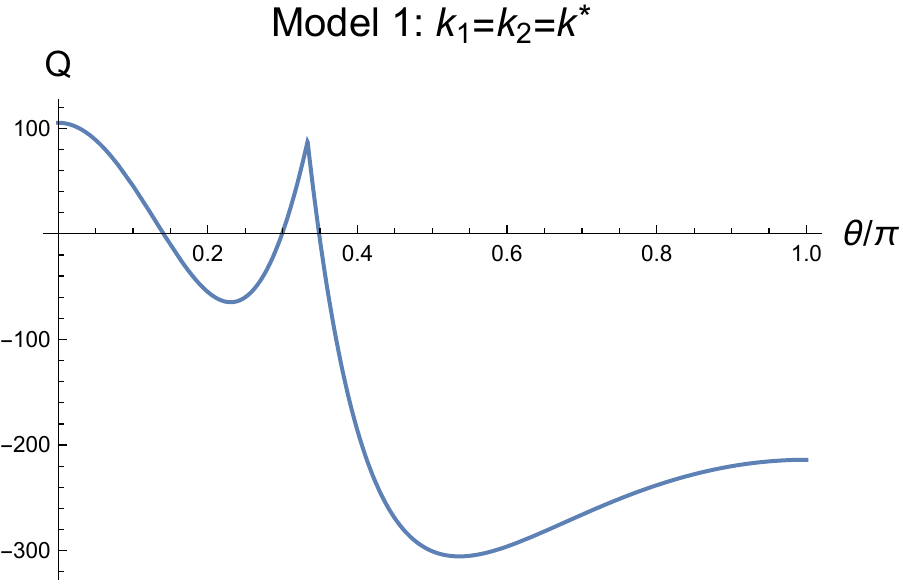}
\includegraphics[width=0.49\linewidth]{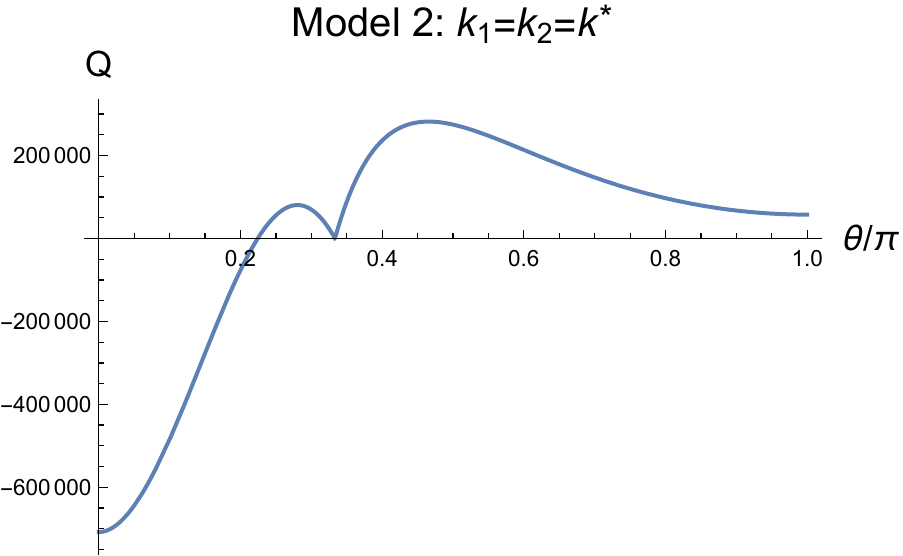} \\
\caption{Reduced bispectra for isosceles triangles with equal sides $k^*$, in terms of the angle between them for the two models. The first model ($r=0.04$) has parameters: $\alpha_B=1$, $\alpha_H=1.04$ and $\beta_K=3.97343$,  $f_2=2.7$, $h_{\mathrm{ds}}=3 \times 10^{-5}$, $f_{0,{\rm xxx}}= 2 \times 10^{-6}$,  $f_{1, {\rm xxx}}=-0.16$, $f_{2,{\rm xxx}}=150$, $f_1=0.0076$, $\beta_B=0$ and $\beta_H=0$. The second model ($r=10^{-3}$) is characterised by: $\alpha_B=1$, $\alpha_H=1.001$, $\beta_K=3.9993$, $h_{\mathrm{ds}}=10^{-5}$, $f_2=8.8$, $f_{0,{\rm xxx}}= 7 \times 10^{-6}$,  $f_{1, {\rm xxx}}=0.12$, $f_{2,{\rm xxx}}=-2675$, $f_1=-0.013$, $\beta_B=0$ and $\beta_H=0.1$.}
\label{fig:qangle}
\end{figure}

In order to investigate the general bispectrum amplitude, we consider the reduced bispectrum, defined as
\begin{align}
Q(k_1,k_2,k_3) = \frac{B(k_1,k_2,k_3)}{P(k_1)P(k_2)+P(k_2)P(k_3)+P(k_3)P(k_1)} \, ,
\end{align}
and we represent it  for isosceles triangles, in terms of the angle between the equal sides \cite{Bartolo:2013ws} (Fig. \ref{fig:qangle}). We note that the amplitude of the reduced bispectrum from the second model is  significantly higher than that of the first model, in line with the results that we found in \cite{Brax:2021qlx}. The plots show that, although  the bispectra are small for particular configurations, the overall bispectrum is large, and hence  potentially detectable. 

The resulting non-Gaussian signal can in principle be large, although its  amplitude and its detectability can only be determined by analysing the \textit{Planck} data. We leave this analysis for future work.

\section{Conclusions}
\label{sec:conclusions}
In this paper we pursue the analysis of DHOST inflation from \cite{Brax:2021qlx}, by studying the bispectrum predicted by such theories using the standard method in the in-in formalism. In pure DHOST models, the bispectrum is scale invariant and  depends on six additional parameters (not present in the power spectrum), which can be tuned such that the local, equilateral and orthogonal \textit{Planck} constraints are satisfied, as well as constraints coming from future CMB surveys. When adding the axion-like perturbations to the potential, the scale invariance is broken, as expected, and the correction terms do not depend on the new parameters. We show that constraints from \textit{Planck} and future probes can be satisfied for the three standard bispectrum templates by fitting the parameters of the models, but the overall bispectrum signal remains large. In a forthcoming paper, we will analyse the \textit{Planck} data and we will make a detailed analysis of the detectability of these models.

\appendix
\section{Operators}
\label{app1}
In this appendix we give the coefficients appearing in the interaction Hamiltonian at cubic order.
\begin{align}
&C_0^{\rm DHOST}=\frac{1}{48 a (\alpha_B+1)^4 H^3} \Large[ 
-432 a^4 H^4 (2 f_2H+f_1) (\alpha_B+1)^4+24 f_2H (k_1^2+k_2^2-k_3^2)^2 \nonumber \\
&\times (\alpha_H+1)^2 (\alpha_B+1)^2+24 f_2H  (k_1^2-k_2^2+k_3^2)^2 (\alpha_H+1)^2 (\alpha_B+1)^2+24 f_2H \nonumber \\
& \times (-k_1^2+k_2^2+k_3^2)^2 (\alpha_H+1)^2 (\alpha_B+1)^2-384 c_s^2 f_2H k_1^2 (k_1^2+k_2^2-k_3^2) \alpha_H (\alpha_H+1) (\alpha_B+1)^2 \nonumber \\
& -384 c_s^2 f_2H k_2^2 (k_1^2+k_2^2-k_3^2) \alpha_H (\alpha_H+1) (\alpha_B+1)^2-384 c_s^2 f_2H k_1^2 (k_1^2-k_2^2+k_3^2) \alpha_H \nonumber \\
& \times(\alpha_H+1) (\alpha_B+1)^2-384 c_s^2 f_2H k_3^2 (k_1^2-k_2^2+k_3^2) \alpha_H (\alpha_H+1) (\alpha_B+1)^2-384 c_s^2 f_2H k_2^2 \nonumber \\
& \times (-k_1^2+k_2^2+k_3^2) \alpha_H (\alpha_H+1) (\alpha_B+1)^2-384 c_s^2 f_2H k_3^2 (-k_1^2+k_2^2+k_3^2) \alpha_H (\alpha_H+1) \nonumber \\
& \times (\alpha_B+1)^2-96 f_2H k_1^2 k_2^2 (\alpha_H+1) (\alpha_H+c_s^2 (2-4 \alpha_H)+1) (\alpha_B+1)^2-96 f_2H k_1^2 k_3^2 \nonumber \\
& \times (\alpha_H+1) (\alpha_H+c_s^2 (2-4 \alpha_H)+1) (\alpha_B+1)^2-96 f_2H k_2^2 k_3^2 (\alpha_H+1) (\alpha_H+c_s^2 (2-4 \alpha_H) \nonumber \\
& +1) (\alpha_B+1)^2-48 a^2 H^2 k_1^2 (f_1 (\alpha_B+1) (3 c_s^2 (\alpha_H-1)-\alpha_H-1)+2 f_2H (3 c_s^2 (\alpha_B^2\nonumber \\
& +(\alpha_H-1) \alpha_B+\alpha_H-\beta_K-2)-2 (\alpha_B+1) (3 \alpha_B+\alpha_H+4))) (\alpha_B+1)^2-48 a^2 H^2 k_2^2 (f_1 
\nonumber \\
& \times (\alpha_B+1) (3 c_s^2 (\alpha_H-1)-\alpha_H-1)+2 f_2H (3 c_s^2 (\alpha_B^2+(\alpha_H-1) \alpha_B+\alpha_H-\beta_K-2)\nonumber \\
& -2 (\alpha_B+1) (3 \alpha_B+\alpha_H+4))) (\alpha_B+1)^2-48 a^2 H^2 k_3^2 (f_1 (\alpha_B+1) (3 c_s^2 (\alpha_H-1)-\alpha_H-1) \nonumber \\
& +2 f_2H (3 c_s^2 (\alpha_B^2+(\alpha_H-1) \alpha_B+\alpha_H-\beta_K-2)-2 (\alpha_B+1) (3 \alpha_B+\alpha_H+4))) (\alpha_B+1)^2
\nonumber \\
& +24 (k_1^2-k_2^2-k_3^2) (k_1^2+k_2^2-k_3^2) (\alpha_H+1)^2 (f_1 (\alpha_H+1)+6 f_2H (\alpha_H-\alpha_B)) (\alpha_B+1) \nonumber \\
& +24 (k_1^2-k_2^2-k_3^2) (k_1^2-k_2^2+k_3^2) (\alpha_H+1)^2 (f_1 (\alpha_H+1)+6 f_2H (\alpha_H-\alpha_B)) (\alpha_B+1) \nonumber \\
& -24 (k_1^2+k_2^2-k_3^2) (k_1^2-k_2^2+k_3^2) (\alpha_H+1)^2 (f_1 (\alpha_H+1)+6 f_2H (\alpha_H-\alpha_B)) (\alpha_B+1) \nonumber \\
& +a H (k_1^2-k_2^2-k_3^2) (24 a H (f_1 (\alpha_H+1) ((\alpha_B+1) (2 \alpha_B+\alpha_H+3)-6 c_s^2 (\alpha_B^2+2 \alpha_B-\beta_K \nonumber \\
&+1))-2 f_2H (-18 \alpha_B^3-2 (3 \alpha_H^2+8 \alpha_H+32) \alpha_B^2-(6 \alpha_H^3+21 \alpha_H^2+32 \alpha_H+71) \alpha_B-6 \alpha_H^3\nonumber \\
&-15 \alpha_H^2-16 \alpha_H+6 c_s^2 (\alpha_B+3 \alpha_H+4) (\alpha_B^2+2 \alpha_B-\beta_K+1)-25))-288 a f_2H^2 (\alpha_B+1) \nonumber \\
& \times (\alpha_H+1)^2 (\alpha_B+\alpha_H)) (\alpha_B+1)-a H (k_1^2+k_2^2-k_3^2) (24 a H (f_1 (\alpha_H+1) ((\alpha_B+1) (2 \alpha_B\nonumber \\
&+\alpha_H+3)-6 c_s^2 (\alpha_B^2+2 \alpha_B-\beta_K+1))-2 f_2H (-18 \alpha_B^3-2 (3 \alpha_H^2+8 \alpha_H+32) \alpha_B^2-(6 \alpha_H^3\nonumber \\
&+21 \alpha_H^2+32 \alpha_H+71) \alpha_B-6 \alpha_H^3-15 \alpha_H^2-16 \alpha_H+6 c_s^2 (\alpha_B+3 \alpha_H+4) (\alpha_B^2+2 \alpha_B-\beta_K\nonumber \\
&+1)-25))-288 a f_2H^2 (\alpha_B+1) (\alpha_H+1)^2 (\alpha_B+\alpha_H)) (\alpha_B+1)-a H (k_1^2-k_2^2+k_3^2) \nonumber \\
& \times (24 a H (f_1 (\alpha_H+1) ((\alpha_B+1) (2 \alpha_B+\alpha_H+3)-6 c_s^2 (\alpha_B^2+2 \alpha_B-\beta_K+1))-2 f_2H \nonumber \\
& \times (-18 \alpha_B^3-2 (3 \alpha_H^2+8 \alpha_H+32) \alpha_B^2-(6 \alpha_H^3+21 \alpha_H^2+32 \alpha_H+71) \alpha_B-6 \alpha_H^3-15 \alpha_H^2 \nonumber \\
&-16 \alpha_H+6 c_s^2 (\alpha_B+3 \alpha_H+4) (\alpha_B^2+2 \alpha_B-\beta_K+1)-25))-288 a f_2H^2 (\alpha_B+1) (\alpha_H+1)^2 \nonumber \\
&\times (\alpha_B+\alpha_H)) (\alpha_B+1)-24 k_1^2 (k_1^2-k_2^2-k_3^2) (\alpha_H+1) (4 f_2H (\alpha_H+1) (\alpha_B+1)^2\nonumber \\
&+c_s^2 (6 f_2H+f_1) (\alpha_H+1) (\alpha_B+1)+3 c_s^2 \alpha_H ((\alpha_B+1) (\alpha_H+1) (f_1+2 f_2H (9-10 \alpha_H))\nonumber \\
&+12 c_s^2 f_2H (1-2 \alpha_H) ((\alpha_B+1)^2-\beta_K))-(\alpha_H+1) ((4 f_2H+f_1) (\alpha_B+1) (\alpha_H+1)\nonumber \\
&-6 c_s^2 f_2H ((\alpha_B+1)^2-\beta_K)))+24 k_3^2 (k_1^2+k_2^2-k_3^2) (\alpha_H+1) (4 f_2H (\alpha_H+1) (\alpha_B+1)^2\nonumber \\
&+c_s^2 (6 f_2H+f_1) (\alpha_H+1) (\alpha_B+1)+3 c_s^2 \alpha_H ((\alpha_B+1) (\alpha_H+1) (f_1+2 f_2H (9-10 \alpha_H))\nonumber \\
&+12 c_s^2 f_2H (1-2 \alpha_H) ((\alpha_B+1)^2-\beta_K))-(\alpha_H+1) ((4 f_2H+f_1) (\alpha_B+1) (\alpha_H+1)\nonumber \\
&-6 c_s^2 f_2H ((\alpha_B+1)^2-\beta_K)))+24 k_2^2 (k_1^2-k_2^2+k_3^2) (\alpha_H+1) (4 f_2H (\alpha_H+1) (\alpha_B+1)^2\nonumber \\
&+c_s^2 (6 f_2H+f_1) (\alpha_H+1) (\alpha_B+1)+3 c_s^2 \alpha_H ((\alpha_B+1) (\alpha_H+1) (f_1+2 f_2H (9-10 \alpha_H))\nonumber \\
&+12 c_s^2 f_2H (1-2 \alpha_H) ((\alpha_B+1)^2-\beta_K))-(\alpha_H+1) ((4 f_2H+f_1) (\alpha_B+1) (\alpha_H+1)\nonumber \\
&-6 c_s^2 f_2H ((\alpha_B+1)^2-\beta_K)))-48 k_1^2 (k_1^2-k_2^2-k_3^2) (\alpha_H+1) (-18 f_2H \alpha_H (2 \alpha_H-1) (\alpha_B^2\nonumber \\
&+2 \alpha_B-\beta_K+1) c_s^4+2 (f_2H (3 \alpha_H^3+(2-3 \beta_K) \alpha_H^2+(3 \beta_K-19) \alpha_H+\alpha_B^2 (6 \alpha_H^2-5 \alpha_H-4)\nonumber \\
&+\alpha_B (3 \alpha_H^3+8 \alpha_H^2-24 \alpha_H-15)+6 \beta_K-11)-f_1 (\alpha_B+1) (\alpha_H+1)^2) c_s^2+(\alpha_B+1) \nonumber \\
& \times (\alpha_H+1) (2 f_2H (\alpha_B-\alpha_H) (\alpha_H-2)+f_1 (\alpha_H+1)))+48 k_3^2 (k_1^2+k_2^2-k_3^2) (\alpha_H+1) \nonumber \\
& \times (-18 f_2H \alpha_H (2 \alpha_H-1) (\alpha_B^2+2 \alpha_B-\beta_K+1) c_s^4+2 (f_2H (3 \alpha_H^3+(2-3 \beta_K) \alpha_H^2+(3 \beta_K\nonumber \\
&-19) \alpha_H+\alpha_B^2 (6 \alpha_H^2-5 \alpha_H-4)+\alpha_B (3 \alpha_H^3+8 \alpha_H^2-24 \alpha_H-15)+6 \beta_K-11)\nonumber \\
&-f_1 (\alpha_B+1) (\alpha_H+1)^2) c_s^2+(\alpha_B+1) (\alpha_H+1) (2 f_2H (\alpha_B-\alpha_H) (\alpha_H-2)+f_1 (\alpha_H+1)))\nonumber \\
&+48 k_2^2 (k_1^2-k_2^2+k_3^2) (\alpha_H+1) (-18 f_2H \alpha_H (2 \alpha_H-1) (\alpha_B^2+2 \alpha_B-\beta_K+1) c_s^4+2 (f_2H \nonumber \\
& \times (3 \alpha_H^3+(2-3 \beta_K) \alpha_H^2+(3 \beta_K-19) \alpha_H+\alpha_B^2 (6 \alpha_H^2-5 \alpha_H-4)+\alpha_B (3 \alpha_H^3+8 \alpha_H^2\nonumber \\
&-24 \alpha_H-15)+6 \beta_K-11)-f_1 (\alpha_B+1) (\alpha_H+1)^2) c_s^2+(\alpha_B+1) (\alpha_H+1) \nonumber \\
& \times (2 f_2H (\alpha_B-\alpha_H) (\alpha_H-2)+f_1 (\alpha_H+1))) 
\Large]
\end{align}

\begin{align}
&C_1^{\rm DHOST}=6 f_2 H a^3+3 f_1 a^3+\frac{9 (2 f_2 H+f_1) \alpha_H a^3}{\alpha_B+1}+\frac{3 (6 f_2 H-8 f_2 H-f_1) (\alpha_B+3 \alpha_H+1) a^3}{\alpha_B+1} \nonumber \\
&-\frac{6 f_2 H \beta_K a^3}{(\alpha_B+1)^2}+\frac{3 f_2 (k_1^2-k_2^2+k_3^2)^2 (\alpha_H+1) (\alpha_B^2+2 \alpha_B-\beta_K+1) a}{2 H k_3^2 (\alpha_B+1)^3}\nonumber \\
&-\frac{3 f_1 \beta_K a^3}{(\alpha_B+1)^2}-\frac{3 (k_1^2-k_2^2+k_3^2) (\alpha_H+1) (6 f_2 H (2 \alpha_B+2 \alpha_H-1)-f_1) (\alpha_B^2+2 \alpha_B-\beta_K+1) a^3}{4 k_3^2 (\alpha_B+1)^3}\nonumber \\
&-\frac{3 (-k_1^2+k_2^2+k_3^2) (\alpha_H+1) (6 f_2 H (2 \alpha_B+2 \alpha_H-1)-f_1) (\alpha_B^2+2 \alpha_B-\beta_K+1) a^3}{4 k_3^2 (\alpha_B+1)^3}\nonumber \\
&+\frac{8 f_2 (k_1^2-k_2^2+k_3^2) \alpha_H a}{\alpha_B H+H}+\frac{3 f_2 (-k_1^2+k_2^2+k_3^2)^2 (\alpha_H+1) (\alpha_B^2+2 \alpha_B-\beta_K+1) a}{2 H k_3^2 (\alpha_B+1)^3}\nonumber \\
&+\frac{3 (k_1^2-k_2^2-k_3^2) (k_1^2+k_2^2-k_3^2) (\alpha_H+1) (f_1 (\alpha_H+1)+6 f_2 H (\alpha_H-\alpha_B)) (\alpha_B^2+2 \alpha_B-\beta_K+1) a}{4 H^2 k_3^2 (\alpha_B+1)^4} \nonumber \\
&-\frac{12 c_s^2 f_2 k_1^2 (k_1^2-k_2^2+k_3^2) \alpha_H (\alpha_B^2+2 \alpha_B-\beta_K+1) a}{H k_3^2 (\alpha_B+1)^3}-\frac{8 f_2 (k_1^2-k_2^2-k_3^2) \alpha_H a}{\alpha_B H+H}\nonumber \\
&-\frac{12 c_s^2 f_2 k_2^2 (-k_1^2+k_2^2+k_3^2) \alpha_H (\alpha_B^2+2 \alpha_B-\beta_K+1) a}{H k_3^2 (\alpha_B+1)^3}\nonumber \\
&-\frac{3 (k_1^2+k_2^2-k_3^2) (k_1^2-k_2^2+k_3^2) (\alpha_H+1) (f_1 (\alpha_H+1)+6 f_2 H (\alpha_H-\alpha_B)) (\alpha_B^2+2 \alpha_B-\beta_K+1) a}{4 H^2 k_3^2 (\alpha_B+1)^4}\nonumber \\
&-\frac{10 f_2 k_3^2 \alpha_H a}{\alpha_B H+H}+\frac{f_2 (k_1^2-k_2^2-k_3^2) (k_1^2-k_2^2+k_3^2) \alpha_H (\alpha_H+1)^2}{a H^3 (\alpha_B+1)^3}\nonumber \\
&+\frac{8 c_s^2 f_2 k_1^2 (k_1^2+k_2^2-k_3^2) \alpha_H (\alpha_H+1) (3 \alpha_H-1)}{a H^3 (\alpha_B+1)^3}+\frac{8 c_s^2 f_2 k_2^2 (k_1^2+k_2^2-k_3^2) \alpha_H (\alpha_H+1) (3 \alpha_H-1)}{a H^3 (\alpha_B+1)^3}\nonumber \\
&+\frac{3 c_s^2 f_2 k_1^2 (k_1^2-k_2^2-k_3^2) \alpha_H (\alpha_H+1)^2 (1-2 \alpha_H)}{a H^3 (\alpha_B+1)^3}+\frac{f_2 (k_1^2+k_2^2-k_3^2)^2 (\alpha_H+1)^2 (1-3 \alpha_H)}{2 a H^3 (\alpha_B+1)^3}\nonumber \\
&+\frac{4 c_s^2 f_2 k_1^2 (k_1^2+k_2^2-k_3^2) \alpha_H (\alpha_H+1) (\alpha_B H+H-3 \alpha_H+1)}{a H^3 (\alpha_B+1)^3}\nonumber \\
&+\frac{4 c_s^2 f_2 k_2^2 (k_1^2+k_2^2-k_3^2) \alpha_H (\alpha_H+1) (\alpha_B H+H-3 \alpha_H+1)}{a H^3 (\alpha_B+1)^3}\nonumber \\
&+\frac{f_2 k_3^2 (k_1^2+k_2^2-k_3^2) \alpha_H (\alpha_H+1)^2 (1-6 \alpha_H)}{a H^3 (\alpha_B+1)^3}\nonumber \\
&+\frac{f_2 (k_1^2+k_2^2-k_3^2) (k_1^2-k_2^2+k_3^2) (\alpha_H+1)^2 (3 \alpha_H^2-\alpha_H+1)}{a H^3 (\alpha_B+1)^3}\nonumber \\
&+\frac{f_2 k_2^2 (k_1^2-k_2^2+k_3^2) (\alpha_H+1) (6 c_s^2 \alpha_H^3+(15 c_s^2+1) \alpha_H^2+(3-7 c_s^2) \alpha_H+2)}{a H^3 (\alpha_B+1)^3}\nonumber \\
&+\frac{3 a (k_1^2-k_2^2-k_3^2) (k_1^2+k_2^2-k_3^2) (\alpha_H+1) (f_1 (\alpha_H+1)-2 f_2 H (\alpha_B-3 \alpha_H-2)) (\alpha_B^2+2 \alpha_B-\beta_K+1)}{4 H^2 k_3^2 (\alpha_B+1)^4}\nonumber \\
&+\frac{3 a (k_1^2+k_2^2-k_3^2) (k_1^2-k_2^2+k_3^2) (\alpha_H+1) (2 f_2 H (\alpha_B-3 \alpha_H-2)-f_1 (\alpha_H+1)) (\alpha_B^2+2 \alpha_B-\beta_K+1)}{4 H^2 k_3^2 (\alpha_B+1)^4}\nonumber \\
&+\frac{3 a (k_1^2-k_2^2-k_3^2) (k_1^2-k_2^2+k_3^2) (\alpha_H+1) (f_1 (\alpha_H+1)+2 f_2 H (-2 \alpha_B+3 \alpha_H+1)) (\alpha_B^2+2 \alpha_B-\beta_K+1)}{2 H^2 k_3^2 (\alpha_B+1)^4}\nonumber \\
&+\frac{1}{16 H k_3^2 (\alpha_B+1)^4}((k_1^2-k_2^2-k_3^2) (24 a^3 H (f_1 (\alpha_B+1) (2 \alpha_B+\alpha_H+3)+2 f_2 H (\alpha_B+1) \nonumber \\
& \times (6 \alpha_H^2+9 \alpha_H+\alpha_B (6 \alpha_H+8)+5)-36 c_s^2 f_2 H (\alpha_B^2+2 \alpha_B-\beta_K+1)-6 c_s^2 f_1 (\alpha_B^2+2 \alpha_B \nonumber \\
&-\beta_K+1))-288 a^3 f_2 H^2 (\alpha_B+1) (\alpha_H+1) (\alpha_B+\alpha_H)) (\alpha_B^2+2 \alpha_B-\beta_K+1))\nonumber \\
&-\frac{1}{16 H k_3^2 (\alpha_B+1)^4} ((k_1^2-k_2^2+k_3^2) (24 a^3 H (f_1 (\alpha_B+1) (2 \alpha_B+\alpha_H+3)+2 f_2 H (\alpha_B+1) \nonumber \\
&\times (6 \alpha_H^2+9 \alpha_H+\alpha_B (6 \alpha_H+8)+5)-36 c_s^2 f_2 H (\alpha_B^2+2 \alpha_B-\beta_K+1)-6 c_s^2 f_1 (\alpha_B^2+2 \alpha_B \nonumber \\
&-\beta_K+1))-288 a^3 f_2 H^2 (\alpha_B+1) (\alpha_H+1) (\alpha_B+\alpha_H)) (\alpha_B^2+2 \alpha_B-\beta_K+1)) \nonumber \\
&-\frac{3 a^3}{4 k_3^2 (\alpha_B+1)^4} ( (k_1^2-k_2^2-k_3^2) (f_1 (\alpha_B+1) (2 \alpha_B+\alpha_H+3)+2 f_2 H (\alpha_B+1) (6 \alpha_H^2+9 \alpha_H \nonumber \\
&+\alpha_B (6 \alpha_H+8)+5)-36 c_s^2 f_2 H (\alpha_B^2+2 \alpha_B-\beta_K+1)-6 c_s^2 f_1 (\alpha_B^2+2 \alpha_B-\beta_K+1)) \nonumber \\
& \times (\alpha_B^2+2 \alpha_B-\beta_K+1)) \nonumber \\
&+\frac{3 a^3}{4 k_3^2 (\alpha_B+1)^4} (k_1^2-k_2^2+k_3^2) (f_1 (\alpha_B+1) (2 \alpha_B+\alpha_H+3)+2 f_2 H (\alpha_B+1) (6 \alpha_H^2+9 \alpha_H \nonumber \\
&+\alpha_B (6 \alpha_H+8)+5)-36 c_s^2 f_2 H (\alpha_B^2+2 \alpha_B-\beta_K+1)-6 c_s^2 f_1 (\alpha_B^2+2 \alpha_B-\beta_K+1)) \nonumber \\
& \times (\alpha_B^2+2 \alpha_B-\beta_K+1)+\frac{a (k_1^2-k_2^2-k_3^2) }{H^2 (\alpha_B+1)^3} (f_1 (\alpha_B+1) (\alpha_H+1)+f_2 H (((3 c_s^2-5) \alpha_H+4) \alpha_B^2 \nonumber \\
&+(13 \alpha_H^2+(6 c_s^2+9) \alpha_H+14) \alpha_B+13 \alpha_H^2+\alpha_H (-3 \beta_K c_s^2+3 c_s^2+14)+10)) \nonumber \\
&-\frac{a (k_1^2-k_2^2+k_3^2) }{H^2 (\alpha_B+1)^3}(f_1 (\alpha_B+1) (\alpha_H+1)+f_2 H (((3 c_s^2-5) \alpha_H+4) \alpha_B^2+(13 \alpha_H^2+(6 c_s^2+9) \alpha_H \nonumber \\
&+14) \alpha_B+13 \alpha_H^2+\alpha_H (-3 \beta_K c_s^2+3 c_s^2+14)+10))-\frac{3 a k_1^2 (k_1^2-k_2^2-k_3^2) }{4 H^2 k_3^2 (\alpha_B+1)^5} \nonumber \\
& \times  (\alpha_B^2+2 \alpha_B-\beta_K+1) (-36 f_2 H \alpha_H (2 \alpha_H-1) (\alpha_B^2+2 \alpha_B-\beta_K+1) c_s^4-(f_1 (\alpha_B+1) \nonumber \\
& \times (\alpha_H^2+4 \alpha_H+3)+18 f_2 H (\alpha_B+1) \alpha_H (2 \alpha_H^2+\alpha_H-1)-2 f_2 H (-18 \alpha_H^3+(4-6 \beta_K) \alpha_H^2 \nonumber \\
&+(3 \beta_K-8) \alpha_H+\alpha_B^2 (12 \alpha_H^2-7 \alpha_H-5)-\alpha_B (18 \alpha_H^3-16 \alpha_H^2+15 \alpha_H+21)+9 \beta_K-16)) c_s^2 \nonumber \\
&+(\alpha_B+1) (\alpha_H+1) (4 f_2 H (\alpha_B-\alpha_H) (\alpha_H-1)+f_1 (\alpha_H+1)))+\frac{3 a k_2^2 (k_1^2-k_2^2+k_3^2) }{4 H^2 k_3^2 (\alpha_B+1)^5} \nonumber \\
&\times (\alpha_B^2+2 \alpha_B-\beta_K+1) (-36 f_2 H \alpha_H (2 \alpha_H-1) (\alpha_B^2+2 \alpha_B-\beta_K+1) c_s^4-(f_1 (\alpha_B+1) \nonumber \\
& \times (\alpha_H^2+4 \alpha_H+3)+18 f_2 H (\alpha_B+1) \alpha_H (2 \alpha_H^2+\alpha_H-1)-2 f_2 H (-18 \alpha_H^3+(4-6 \beta_K) \alpha_H^2 \nonumber \\
& +(3 \beta_K-8) \alpha_H+\alpha_B^2 (12 \alpha_H^2-7 \alpha_H-5)-\alpha_B (18 \alpha_H^3-16 \alpha_H^2+15 \alpha_H+21)+9 \beta_K-16)) c_s^2 \nonumber \\
&+(\alpha_B+1) (\alpha_H+1) (4 f_2 H (\alpha_B-\alpha_H) (\alpha_H-1)+f_1 (\alpha_H+1)))+\frac{3 a k_1^2 (k_1^2-k_2^2-k_3^2) }{4 H^2 k_3^2 (\alpha_B+1)^5} \nonumber \\
&\times (\alpha_B^2+2 \alpha_B-\beta_K+1) (f_1 (\alpha_B+1) (\alpha_H+1) (-\alpha_H+c_s^2 (\alpha_H+3)-1)+2 f_2 (9 c_s^2 H (\alpha_B+1) \nonumber \\
& \times \alpha_H (2 \alpha_H^2+\alpha_H-1)+H (18 \alpha_H (2 \alpha_H-1) (\alpha_B^2+2 \alpha_B-\beta_K+1) c_s^4+(\alpha_H+1) ((1-12 \alpha_H) \alpha_B^2 \nonumber \\
&+(18 \alpha_H^2-34 \alpha_H+13) \alpha_B+18 \alpha_H^2-22 \alpha_H+6 \alpha_H \beta_K-3 \beta_K+12) c_s^2+2 (\alpha_H^3+\alpha_H^2+\alpha_H \nonumber \\
&-\alpha_B^2 (\alpha_H^2-1)+\alpha_B (\alpha_H^3+\alpha_H+2)+1))))-\frac{3 a k_2^2 (k_1^2-k_2^2+k_3^2) (\alpha_B^2+2 \alpha_B-\beta_K+1) }{4 H^2 k_3^2 (\alpha_B+1)^5} \nonumber \\
& \times (f_1 (\alpha_B+1) (\alpha_H+1) (-\alpha_H+c_s^2 (\alpha_H+3)-1)+2 f_2 (9 c_s^2 H (\alpha_B+1) \alpha_H (2 \alpha_H^2+\alpha_H-1) \nonumber \\
&+H (18 \alpha_H (2 \alpha_H-1) (\alpha_B^2+2 \alpha_B-\beta_K+1) c_s^4+(\alpha_H+1) ((1-12 \alpha_H) \alpha_B^2+(18 \alpha_H^2-34 \alpha_H \nonumber \\
&+13) \alpha_B+18 \alpha_H^2-22 \alpha_H+6 \alpha_H \beta_K-3 \beta_K+12) c_s^2+2 (\alpha_H^3+\alpha_H^2+\alpha_H-\alpha_B^2 (\alpha_H^2-1)\nonumber \\ 
&+\alpha_B (\alpha_H^3+\alpha_H+2)+1))))+\frac{a k_1^2 }{H^2 (\alpha_B+1)^3} (f_1 (\alpha_B+1) (3 c_s^2 (\alpha_H-3)-\alpha_H+1) (\alpha_H+1) \nonumber \\
&+2 f_2 (H (-((13 \alpha_H+5) \alpha_B^2)+(2 \alpha_H^2-23 \alpha_H-9) \alpha_B+2 \alpha_H^2-10 \alpha_H+3 \alpha_H \beta_K+3 \beta_K \nonumber \\
&+3 c_s^2 (11 \alpha_H \alpha_B^2+(10 \alpha_H^2+17 \alpha_H-11) \alpha_B+10 \alpha_H^2+6 \alpha_H-5 \alpha_H \beta_K+2 \beta_K-11)-4) \nonumber \\
&-9 c_s^2 H (\alpha_B+1) \alpha_H (3 \alpha_H-1))) +\frac{a k_2^2 }{H^2 (\alpha_B+1)^3} (f_1 (\alpha_B+1) (3 c_s^2 (\alpha_H-3)-\alpha_H+1) \nonumber \\
& \times (\alpha_H+1)+2 f_2 (H (-((13 \alpha_H+5) \alpha_B^2)+(2 \alpha_H^2-23 \alpha_H-9) \alpha_B+2 \alpha_H^2-10 \alpha_H+3 \alpha_H \beta_K \nonumber \\
&+3 \beta_K+3 c_s^2 (11 \alpha_H \alpha_B^2+(10 \alpha_H^2+17 \alpha_H-11) \alpha_B+10 \alpha_H^2+6 \alpha_H-5 \alpha_H \beta_K+2 \beta_K-11)-4) \nonumber \\
&-9 c_s^2 H (\alpha_B+1) \alpha_H (3 \alpha_H-1))) +\frac{(k_1^2+k_2^2-k_3^2)}{48 H^3 (\alpha_B+1)^5} (288 a f_2 H^2 (\alpha_B+1) (\alpha_H+1)^2 ((5 \alpha_H-4) \alpha_B^2 \nonumber \\
&+(8 \alpha_H^2+5 \alpha_H-\beta_B+4 \beta_H+\beta_K-4) \alpha_B+8 \alpha_H^2-\beta_B+4 \beta_H)-24 a H (f_1 (\alpha_B+1) (\alpha_H+1) \nonumber \\
& \times (-4 \alpha_H \alpha_B^2+2 (\alpha_H^2+3) \alpha_B+2 \alpha_H^2+4 \alpha_H+6 c_s^2 (\alpha_H+3) (\alpha_B^2+2 \alpha_B-\beta_K+1)+3 \alpha_H \beta_K \nonumber \\
& +3 \beta_K+6)+2 f_2 (9 H (\alpha_B+1)^2 (2 \alpha_H-1) (\alpha_B+3 \alpha_H+1) (\alpha_H+1)^2-H (\alpha_B+1) (-66 \alpha_H^4 \nonumber \\
&-115 \alpha_H^3+3 (2 \beta_B-8 \beta_H-\beta_K-25) \alpha_H^2+3 (4 \beta_B-16 \beta_H-2 \beta_K-7) \alpha_H +\alpha_B^3 (22 \alpha_H-2)\nonumber \\
&+\alpha_B^2 (-24 \alpha_H^3-45 \alpha_H^2+74 \alpha_H+23)+6 \beta_B-24 \beta_H-3 \beta_K-\alpha_B (66 \alpha_H^4+139 \alpha_H^3-6 (\beta_B \nonumber \\
&-4 \beta_H-\beta_K-20) \alpha_H^2+(-12 \beta_B+48 \beta_H+12 \beta_K-31) \alpha_H+6 (-\beta_B+4 \beta_H+\beta_K-1))-19) \nonumber \\
&+6 c_s^2 H ((6 \alpha_H^2+4 \alpha_H-6) \alpha_B^4+(36 \alpha_H^3+41 \alpha_H^2+8 \alpha_H-13) \alpha_B^3+(108 \alpha_H^3+(87-6 \beta_K) \alpha_H^2 \nonumber \\
&-7 \beta_K \alpha_H+3 (\beta_K-1)) \alpha_B^2-(36 (\beta_K-3) \alpha_H^3+(29 \beta_K-75) \alpha_H^2+(6 \beta_K+8) \alpha_H+5 \beta_K-9) \alpha_B \nonumber \\
&+3 \beta_K^2-36 \alpha_H^3 (\beta_K-1)-23 \alpha_H^2 (\beta_K-1)-8 \beta_K+\alpha_H (3 \beta_K^2+\beta_K-4)+5)))) \nonumber \\
&-\frac{3 c_s^2 f_2 k_2^2 (k_1^2-k_2^2+k_3^2) (1-2 \alpha_H) \alpha_H (\alpha_H+1)^2}{a H^3 (\alpha_B+1)^3} \nonumber \\
&-\frac{f_2 (k_1^2-k_2^2-k_3^2) (k_1^2+k_2^2-k_3^2) (\alpha_H+1)^2 (3 \alpha_H^2-\alpha_H+1)}{a H^3 (\alpha_B+1)^3} \nonumber \\
&-\frac{f_2 k_1^2 (k_1^2-k_2^2-k_3^2) (\alpha_H+1) (6 c_s^2 \alpha_H^3+(15 c_s^2+1) \alpha_H^2+(3-7 c_s^2) \alpha_H+2)}{a H^3 (\alpha_B+1)^3} \nonumber \\
&-\frac{2 f_2 k_1^2 k_2^2 (\alpha_H+1) (-3 \alpha_H^2-2 \alpha_H+2 c_s^2 (6 \alpha_H^2-\alpha_H-3)+1)}{a H^3 (\alpha_B+1)^3}
\end{align}

\begin{align}
&C_2^{\rm DHOST}=\frac{a}{4 (\alpha_B+1)^6 H^3 k_2^2 k_3^2} \Big[ -18 a^4 (k_1^2-k_2^2-k_3^2) (6 f_2H+f_1) (\alpha_B+1)^2 (\alpha_B^2+2 \alpha_B \nonumber \\
& -\beta_K+1)^2 H^4+18 a^2 f_2(-k_1^2+k_2^2+k_3^2)^2 (\alpha_B+1)^2 (\alpha_B^2+2 \alpha_B-\beta_K+1)^2 H^3+36 a^2 f_2k_3^2 \nonumber \\
& \times (k_1^2+k_2^2-k_3^2) (\alpha_B+1) (\alpha_H+1) (\alpha_B^2+2 \alpha_B-\beta_K+1) ((5 \alpha_H-4) \alpha_B^2+(8 \alpha_H^2+5 \alpha_H-\beta_B \nonumber \\
&+4 \beta_H+\beta_K-4) \alpha_B+8 \alpha_H^2-\beta_B+4 \beta_H) H^3+36 a^2 f_2k_2^2 (k_1^2-k_2^2+k_3^2) (\alpha_B+1) (\alpha_H+1) \nonumber \\
& \times (\alpha_B^2+2 \alpha_B-\beta_K+1) ((5 \alpha_H-4) \alpha_B^2+(8 \alpha_H^2+5 \alpha_H-\beta_B+4 \beta_H+\beta_K-4) \alpha_B+8 \alpha_H^2\nonumber \\
&-\beta_B+4 \beta_H) H^3+72 a^2 f_2k_1^2 (k_1^2-k_2^2-k_3^2) (\alpha_B^2+2 \alpha_B-\beta_K+1)^2 (((\alpha_H-1)^2-c_s^2 (9 \alpha_H^2\nonumber \\
&+11 \alpha_H+1)) \alpha_B-\alpha_B^2 (6 \alpha_H c_s^2+c_s^2+\alpha_H-1)+\alpha_H (\alpha_H+c_s^2 (-9 \alpha_H+3 \beta_K-5)-1)) H^3\nonumber \\
&+9 a^2 (k_1^2-k_2^2-k_3^2) (k_1^2+k_2^2-k_3^2) (6 f_2H+f_1) (\alpha_B+1) (\alpha_H+1) (\alpha_B^2+2 \alpha_B-\beta_K+1)^2 H^2\nonumber \\
&+9 a^2 (k_1^2-k_2^2-k_3^2) (k_1^2-k_2^2+k_3^2) (6 f_2H+f_1) (\alpha_B+1) (\alpha_H+1) (\alpha_B^2+2 \alpha_B-\beta_K+1)^2 H^2\nonumber \\
&+18 a^2 (k_1^2+k_2^2-k_3^2) (k_1^2-k_2^2+k_3^2) (\alpha_B+1) (2 f_2H (\alpha_B-3 \alpha_H-2)-f_1 (\alpha_H+1)) (\alpha_B^2+2 \alpha_B\nonumber \\
&-\beta_K+1)^2 H^2-9 a^2 (k_1^2-k_2^2-k_3^2) (k_1^2+k_2^2-k_3^2) (\alpha_B+1) (2 f_2H (2 \alpha_B-3 \alpha_H-1)\nonumber \\
&-f_1 (\alpha_H+1)) (\alpha_B^2+2 \alpha_B-\beta_K+1)^2 H^2-9 a^2 (k_1^2-k_2^2-k_3^2) (k_1^2-k_2^2+k_3^2) (\alpha_B+1) (2 f_2H \nonumber \\
&\times  (2 \alpha_B-3 \alpha_H-1)-f_1 (\alpha_H+1)) (\alpha_B^2+2 \alpha_B-\beta_K+1)^2 H^2+18 a^2 k_1^2 (k_1^2-k_2^2-k_3^2) \nonumber \\
& \times (f_1 (\alpha_B+1) (-\alpha_H+c_s^2 (\alpha_H+3)-1)+2 f_2(2 H (\alpha_B+1) (\alpha_H+1)+9 c_s^2 H (\alpha_B+1) \alpha_H \nonumber \\
& \times (2 \alpha_H-1)+3 c_s^2 H (\alpha_B^2+(16 \alpha_H^2-6 \alpha_H+5) \alpha_B+16 \alpha_H^2-6 \alpha_H-\beta_K+4))) (\alpha_B^2+2 \alpha_B\nonumber \\
&-\beta_K+1)^2 H^2+12 a^2 k_2^2 (k_1^2-k_2^2-k_3^2) (\alpha_B+1)^3 (f_1+2 f_2H (2 \alpha_B+7 \alpha_H+5)) (\alpha_B^2+2 \alpha_B\nonumber \\
&-\beta_K+1) H^2+12 a^2 k_3^2 (k_1^2-k_2^2-k_3^2) (\alpha_B+1)^3 (f_1+2 f_2H (2 \alpha_B+7 \alpha_H+5)) (\alpha_B^2+2 \alpha_B\nonumber \\
&-\beta_K+1) H^2-3 a^2 k_3^2 (k_1^2+k_2^2-k_3^2) (\alpha_B+1) (\alpha_B^2+2 \alpha_B-\beta_K+1) (f_1 (\alpha_H+1) (-2 \alpha_B^2\nonumber \\
&+4 (\alpha_H+2) \alpha_B+4 \alpha_H+3 \beta_K+10)+2 f_2(9 H (\alpha_B+1) (\alpha_B+3 \alpha_H+1) (2 \alpha_H^2+\alpha_H-1)\nonumber \\
&+18 c_s^2 H (2 \alpha_H-1) (\alpha_B+3 \alpha_H+1) (\alpha_B^2+2 \alpha_B-\beta_K+1)+H (\alpha_H+1) ((24 \alpha_H-35) \alpha_B^2\nonumber \\
&+(102 \alpha_H^2-13 \alpha_H-6 \beta_B+24 \beta_H+6 \beta_K-2) \alpha_B+102 \alpha_H^2-37 \alpha_H-6 \beta_B+24 \beta_H-3 \beta_K\nonumber \\
&+33))) H^2-3 a^2 k_2^2 (k_1^2-k_2^2+k_3^2) (\alpha_B+1) (\alpha_B^2+2 \alpha_B-\beta_K+1) (f_1 (\alpha_H+1) (-2 \alpha_B^2\nonumber \\
&+4 (\alpha_H+2) \alpha_B+4 \alpha_H+3 \beta_K+10)+2 f_2(9 H (\alpha_B+1) (\alpha_B+3 \alpha_H+1) (2 \alpha_H^2+\alpha_H-1)\nonumber \\
&+18 c_s^2 H (2 \alpha_H-1) (\alpha_B+3 \alpha_H+1) (\alpha_B^2+2 \alpha_B-\beta_K+1)+H (\alpha_H+1) ((24 \alpha_H-35) \alpha_B^2\nonumber \\
&+(102 \alpha_H^2-13 \alpha_H-6 \beta_B+24 \beta_H+6 \beta_K-2) \alpha_B+102 \alpha_H^2-37 \alpha_H-6 \beta_B+24 \beta_H\nonumber \\
&-3 \beta_K+33))) H^2+12 a^2 k_2^2 k_3^2 (\alpha_B+1)^2 (f_1 (\alpha_B+1) (3 \alpha_H^2+2 (\beta_K-3) \alpha_H+3 \alpha_B (\alpha_H^2-2 \alpha_H\nonumber \\
&-3)-2 \beta_K-9)+2 f_2(H (3 \alpha_B^4+6 (8 \alpha_H+1) \alpha_B^3+3 (19 \alpha_H^2+40 \alpha_H-11) \alpha_B^2+2 (57 \alpha_H^2\nonumber \\
&-5 \beta_K \alpha_H+48 \alpha_H-36) \alpha_B+57 \alpha_H^2-3 \beta_K^2+24 \alpha_H-10 \alpha_H \beta_K-36)-6 H (\alpha_B+1)^2 \nonumber \\
& \times (3 \alpha_H-1) (\alpha_B+3 \alpha_H+1))) H^2+3 a^2 k_3^2 (k_1^2+k_2^2-k_3^2) (\alpha_B^2+2 \alpha_B-\beta_K+1) (-f_1 (\alpha_B+1) \nonumber \\
& \times (-4 \alpha_H \alpha_B^2+2 (\alpha_H^2+3) \alpha_B+2 \alpha_H^2+4 \alpha_H+6 c_s^2 (\alpha_H+3) (\alpha_B^2+2 \alpha_B-\beta_K+1)+3 \alpha_H \beta_K\nonumber \\
&+3 \beta_K+6)+12 f_2H (\alpha_B+1) (\alpha_H+1) ((5 \alpha_H-4) \alpha_B^2+(8 \alpha_H^2+5 \alpha_H-\beta_B+4 \beta_H+\beta_K-4) \alpha_B\nonumber \\
&+8 \alpha_H^2-\beta_B+4 \beta_H)-2 f_2(9 H (\alpha_B+3 \alpha_H+1) (2 \alpha_H^2+\alpha_H-1) (\alpha_B+1)^2+H (8 \alpha_B^3+(36 \alpha_H^2\nonumber \\
&+33 \alpha_H-13) \alpha_B^2+(102 \alpha_H^3+115 \alpha_H^2+(-6 \beta_B+24 \beta_H+6 \beta_K+53) \alpha_H-6 \beta_B+24 \beta_H+6 \beta_K\nonumber \\
&-4) \alpha_B+102 \alpha_H^3+79 \alpha_H^2-6 \beta_B+24 \beta_H+3 \beta_K+\alpha_H (-6 \beta_B+24 \beta_H+3 \beta_K+20)+17) \nonumber \\
& \times (\alpha_B+1)-6 c_s^2 H (\alpha_B^2+2 \alpha_B-\beta_K+1) (2 \alpha_B^2+(-18 \alpha_H^2+10 \alpha_H-7) \alpha_B-18 \alpha_H^2+10 \alpha_H\nonumber \\
&+3 \beta_K-9))) H^2+3 a^2 k_2^2 (k_1^2-k_2^2+k_3^2) (\alpha_B^2+2 \alpha_B-\beta_K+1) (-f_1 (\alpha_B+1) (-4 \alpha_H \alpha_B^2\nonumber \\
&+2 (\alpha_H^2+3) \alpha_B+2 \alpha_H^2+4 \alpha_H+6 c_s^2 (\alpha_H+3) (\alpha_B^2+2 \alpha_B-\beta_K+1)+3 \alpha_H \beta_K+3 \beta_K+6)\nonumber \\
&+12 f_2H (\alpha_B+1) (\alpha_H+1) ((5 \alpha_H-4) \alpha_B^2+(8 \alpha_H^2+5 \alpha_H-\beta_B+4 \beta_H+\beta_K-4) \alpha_B+8 \alpha_H^2\nonumber \\
&-\beta_B+4 \beta_H)-2 f_2(9 H (\alpha_B+3 \alpha_H+1) (2 \alpha_H^2+\alpha_H-1) (\alpha_B+1)^2+H (8 \alpha_B^3+(36 \alpha_H^2+33 \alpha_H\nonumber \\
&-13) \alpha_B^2+(102 \alpha_H^3+115 \alpha_H^2+(-6 \beta_B+24 \beta_H+6 \beta_K+53) \alpha_H-6 \beta_B+24 \beta_H+6 \beta_K-4) \alpha_B\nonumber \\
&+102 \alpha_H^3+79 \alpha_H^2-6 \beta_B+24 \beta_H+3 \beta_K+\alpha_H (-6 \beta_B+24 \beta_H+3 \beta_K+20)+17) (\alpha_B+1)\nonumber \\
&-6 c_s^2 H (\alpha_B^2+2 \alpha_B-\beta_K+1) (2 \alpha_B^2+(-18 \alpha_H^2+10 \alpha_H-7) \alpha_B-18 \alpha_H^2+10 \alpha_H+3 \beta_K-9))) H^2\nonumber \\
&+8 f_2k_2^2 k_3^4 (\alpha_B+1)^4 \alpha_H (7 \alpha_H+3) H+8 f_2k_2^4 k_3^2 (\alpha_B+1)^4 \alpha_H (7 \alpha_H+3) H-8 f_2k_2^2 k_3^2 (-k_1^2+k_2^2\nonumber \\
&+k_3^2) (\alpha_B+1)^4 (5 \alpha_H^2-1) H+4 f_2k_2^2 k_3^2 (k_1^2+k_2^2-k_3^2) (\alpha_B+1)^3 (18 \alpha_H^4+33 \alpha_H^3-3 \alpha_H^2-13 \alpha_H\nonumber \\
&+\alpha_B (6 \alpha_H^3-3 \alpha_H^2-4 \alpha_H-3)-3) H-4 f_2k_2^2 k_3^2 (k_1^2-k_2^2+k_3^2) (\alpha_B+1)^3 (-18 \alpha_H^4-33 \alpha_H^3\nonumber \\
&+3 \alpha_H^2+13 \alpha_H+\alpha_B (-6 \alpha_H^3+3 \alpha_H^2+4 \alpha_H+3)+3) H-12 f_2k_2^2 (k_1^2+k_2^2-k_3^2) \nonumber \\
& \times  (-k_1^2+k_2^2+k_3^2) (\alpha_B+1)^2 \alpha_H (\alpha_H+1) (\alpha_B^2+2 \alpha_B-\beta_K+1) H-12 f_2k_3^2 (k_1^2-k_2^2+k_3^2) \nonumber \\
& \times (-k_1^2+k_2^2+k_3^2) (\alpha_B+1)^2 \alpha_H (\alpha_H+1) (\alpha_B^2+2 \alpha_B-\beta_K+1) H+48 c_s^2 f_2k_1^2 k_3^2 (k_1^2+k_2^2-k_3^2) \nonumber \\
& \times (\alpha_B+1)^2 \alpha_H (3 \alpha_H-1) (\alpha_B^2+2 \alpha_B-\beta_K+1) H+48 c_s^2 f_2k_1^2 k_2^2 (k_1^2-k_2^2+k_3^2) (\alpha_B+1)^2 \alpha_H \nonumber \\
& \times  (3 \alpha_H-1) (\alpha_B^2+2 \alpha_B-\beta_K+1) H+3 f_2k_3^2 (k_1^2+k_2^2-k_3^2)^2 (\alpha_B+1)^2 (\alpha_H+1) (1-3 \alpha_H) \nonumber \\
& \times (\alpha_B^2+2 \alpha_B-\beta_K+1) H+3 f_2k_2^2 (k_1^2-k_2^2+k_3^2)^2 (\alpha_B+1)^2 (\alpha_H+1) (1-3 \alpha_H) (\alpha_B^2+2 \alpha_B\nonumber \\
&-\beta_K+1) H+6 f_2k_3^4 (k_1^2+k_2^2-k_3^2) (\alpha_B+1)^2 \alpha_H (\alpha_H+1) (1-6 \alpha_H) (\alpha_B^2+2 \alpha_B-\beta_K+1) H\nonumber \\
&+6 f_2k_2^4 (k_1^2-k_2^2+k_3^2) (\alpha_B+1)^2 \alpha_H (\alpha_H+1) (1-6 \alpha_H) (\alpha_B^2+2 \alpha_B-\beta_K+1) H+72 c_s^2 f_2k_2^2 k_3^2 \nonumber \\
& \times (-k_1^2+k_2^2+k_3^2) (\alpha_B+1)^2 \alpha_H (2 \alpha_H^2+\alpha_H-1) (\alpha_B^2+2 \alpha_B-\beta_K+1) H+12 f_2k_2^2 \nonumber \\
& \times (k_1^2+k_2^2-k_3^2) (k_1^2-k_2^2+k_3^2) (\alpha_B+1)^2 (\alpha_H+1) (3 \alpha_H^2-\alpha_H+1) (\alpha_B^2+2 \alpha_B-\beta_K+1) H\nonumber \\
&+12 f_2k_3^2 (k_1^2+k_2^2-k_3^2) (k_1^2-k_2^2+k_3^2) (\alpha_B+1)^2 (\alpha_H+1) (3 \alpha_H^2-\alpha_H+1) (\alpha_B^2+2 \alpha_B-\beta_K\nonumber \\
&+1) H-12 f_2k_2^2 (-k_1^2+k_2^2-k_3^2) (-k_1^2+k_2^2+k_3^2) (\alpha_B+1)^2 (\alpha_H+1) (3 \alpha_H^2-\alpha_H+1) (\alpha_B^2\nonumber \\
&+2 \alpha_B-\beta_K+1) H-12 f_2k_3^2 (-k_1^2-k_2^2+k_3^2) (-k_1^2+k_2^2+k_3^2) (\alpha_B+1)^2 (\alpha_H+1) (3 \alpha_H^2-\alpha_H\nonumber \\
&+1) (\alpha_B^2+2 \alpha_B-\beta_K+1) H+3 f_2k_3^2 (k_1^2+k_2^2-k_3^2)^2 (\alpha_B+1)^2 (-3 \alpha_H^2-2 \alpha_H+1) (\alpha_B^2+2 \alpha_B\nonumber \\
&-\beta_K+1) H+3 f_2k_2^2 (k_1^2-k_2^2+k_3^2)^2 (\alpha_B+1)^2 (-3 \alpha_H^2-2 \alpha_H+1) (\alpha_B^2+2 \alpha_B-\beta_K+1) H\nonumber \\
&+6 f_2k_3^4 (k_1^2+k_2^2-k_3^2) (\alpha_B+1)^2 \alpha_H (-6 \alpha_H^2-5 \alpha_H+1) (\alpha_B^2+2 \alpha_B-\beta_K+1) H+6 f_2k_2^4 \nonumber \\
& \times (k_1^2-k_2^2+k_3^2) (\alpha_B+1)^2 \alpha_H (-6 \alpha_H^2-5 \alpha_H+1) (\alpha_B^2+2 \alpha_B-\beta_K+1) H-12 f_2k_1^2 k_2^2 (k_1^2\nonumber \\
&-k_2^2-k_3^2) (\alpha_B+1)^2 (6 c_s^2 \alpha_H^3+(15 c_s^2+1) \alpha_H^2+(3-7 c_s^2) \alpha_H+2) (\alpha_B^2+2 \alpha_B-\beta_K+1) H\nonumber \\
&-12 f_2k_1^2 k_3^2 (k_1^2-k_2^2-k_3^2) (\alpha_B+1)^2 (6 c_s^2 \alpha_H^3+(15 c_s^2+1) \alpha_H^2+(3-7 c_s^2) \alpha_H+2) (\alpha_B^2+2 \alpha_B\nonumber \\
&-\beta_K+1) H+4 k_2^2 k_3^2 (k_1^2+k_2^2-k_3^2) (\alpha_B+1)^2 (3 f_1 (\alpha_B+1) (\alpha_H+1)^2+f_2H (18 \alpha_H^4+94 \alpha_H^3\nonumber \\
&+(-9 c_s^2 (\beta_K-1)-3 \beta_K+55) \alpha_H^2+(3 c_s^2 (\beta_K-1)-9 \beta_K+5) \alpha_H+\alpha_B^2 ((9 c_s^2-1) \alpha_H^2\nonumber \\
&-3 (c_s^2-1) \alpha_H-8)+2 \alpha_B (9 \alpha_H^4+47 \alpha_H^3+9 (c_s^2+3) \alpha_H^2+(4-3 c_s^2) \alpha_H+3)-6 \beta_K+14))\nonumber \\
&+4 k_2^2 k_3^2 (k_1^2-k_2^2+k_3^2) (\alpha_B+1)^2 (3 f_1 (\alpha_B+1) (\alpha_H+1)^2+f_2H (18 \alpha_H^4+94 \alpha_H^3\nonumber \\
&+(-9 c_s^2 (\beta_K-1)-3 \beta_K+55) \alpha_H^2+(3 c_s^2 (\beta_K-1)-9 \beta_K+5) \alpha_H+\alpha_B^2 ((9 c_s^2-1) \alpha_H^2\nonumber \\
&-3 (c_s^2-1) \alpha_H-8)+2 \alpha_B (9 \alpha_H^4+47 \alpha_H^3+9 (c_s^2+3) \alpha_H^2+(4-3 c_s^2) \alpha_H+3)-6 \beta_K+14))\nonumber \\
&-4 k_1^2 k_2^2 k_3^2 (\alpha_B+1)^2 (f_1 (\alpha_B+1) (3 c_s^2 (\alpha_H-5)-\alpha_H+3) (\alpha_H+1)^2+2 f_2(-9 c_s^2 H (\alpha_B+1) \nonumber \\
& \times \alpha_H (5 \alpha_H^2+10 \alpha_H-4 \beta_H-3)-2 H (-15 \alpha_H^3+(2-9 \beta_K) \alpha_H^2+(\beta_B-4 \beta_H-6 \beta_K+13) \alpha_H\nonumber \\
&+\beta_B+4 \alpha_B^2 (3 \alpha_H^2+2 \alpha_H+\beta_H+1)+\alpha_B (-15 \alpha_H^3+14 \alpha_H^2+(\beta_B-4 \beta_H+21) \alpha_H+\beta_B\nonumber \\
&+4 \beta_H+8)+3 \beta_K+4)+c_s^2 H (48 \alpha_H^3+(198-51 \beta_K) \alpha_H^2+(6 \beta_B-48 \beta_H-24 \beta_K+22) \alpha_H\nonumber \\
&+2 \beta_B-48 \beta_H+\alpha_B^2 (111 \alpha_H^2+120 \alpha_H-36 \beta_H-7)+\alpha_B (48 \alpha_H^3+309 \alpha_H^2+2 (3 \beta_B-24 \beta_H\nonumber \\
&+71) \alpha_H+2 \beta_B-84 \beta_H-79)+12 \beta_H \beta_K+27 \beta_K-72))) \Big]
\end{align}

\begin{align}
&C_3^{\rm DHOST}=\frac{a}{2 (\alpha_B+1)^6 H^3 k_1^2 k_2^2 k_3^2} \Big[ 27 H^3 (k_1^2-k_2^2-k_3^2) (k_1^2+k_2^2-k_3^2) (6 f_2 H+f_1) (\alpha_B^2+2 \alpha_B \nonumber \\
&-\beta_K+1)^3 a^4+27 H^3 (k_1^2-k_2^2-k_3^2) (k_1^2-k_2^2+k_3^2) (6 f_2 H+f_1) (\alpha_B^2+2 \alpha_B-\beta_K+1)^3 a^4\nonumber \\
&-27 H^3 (k_1^2+k_2^2-k_3^2) (k_1^2-k_2^2+k_3^2) (6 f_2 H+f_1) (\alpha_B^2+2 \alpha_B-\beta_K+1)^3 a^4+18 H^3 k_1^2 (k_1^2\nonumber \\
&-k_2^2-k_3^2) (\alpha_B^2+2 \alpha_B-\beta_K+1)^2 (f_1 (-\alpha_B^2+5 \alpha_H \alpha_B+\alpha_B+5 \alpha_H+3 \beta_K+2)+2 f_2 H (8 \alpha_B^2\nonumber \\
&+(36 \alpha_H^2+7 \alpha_H+32) \alpha_B+36 \alpha_H^2+7 \alpha_H-3 \beta_K+24)) a^4-18 H^3 k_3^2 (k_1^2+k_2^2-k_3^2) (\alpha_B^2+2 \alpha_B\nonumber \\
&-\beta_K+1)^2 (f_1 (-\alpha_B^2+5 \alpha_H \alpha_B+\alpha_B+5 \alpha_H+3 \beta_K+2)+2 f_2 H (8 \alpha_B^2+(36 \alpha_H^2+7 \alpha_H+32) \alpha_B\nonumber \\
&+36 \alpha_H^2+7 \alpha_H-3 \beta_K+24)) a^4-18 H^3 k_2^2 (k_1^2-k_2^2+k_3^2) (\alpha_B^2+2 \alpha_B-\beta_K+1)^2 (f_1 (-\alpha_B^2\nonumber \\
&+5 \alpha_H \alpha_B+\alpha_B+5 \alpha_H+3 \beta_K+2)+2 f_2 H (8 \alpha_B^2+(36 \alpha_H^2+7 \alpha_H+32) \alpha_B+36 \alpha_H^2+7 \alpha_H\nonumber \\
&-3 \beta_K+24)) a^4+27 H^3 k_1^2 (k_1^2-k_2^2-k_3^2) (\alpha_B^2+2 \alpha_B-\beta_K+1)^2 (2 f_2 (3 H (\alpha_B+1) (2 \alpha_H-1) \nonumber \\
& \times (\alpha_B+3 \alpha_H+1)+H ((2 \alpha_H-11) \alpha_B^2+(6 \alpha_H^2-17 \alpha_H-18) \alpha_B+6 \alpha_H^2-19 \alpha_H+\beta_K-7))\nonumber \\
&-f_1 (2 \alpha_B (\alpha_H-1)+2 \alpha_H+\beta_K-2)) a^4-27 H^3 k_3^2 (k_1^2+k_2^2-k_3^2) (\alpha_B^2+2 \alpha_B-\beta_K+1)^2 (2 f_2 \nonumber \\
& \times (3 H (\alpha_B+1) (2 \alpha_H-1) (\alpha_B+3 \alpha_H+1)+H ((2 \alpha_H-11) \alpha_B^2+(6 \alpha_H^2-17 \alpha_H-18) \alpha_B\nonumber \\
&+6 \alpha_H^2-19 \alpha_H+\beta_K-7))-f_1 (2 \alpha_B (\alpha_H-1)+2 \alpha_H+\beta_K-2)) a^4-27 H^3 k_2^2 (k_1^2-k_2^2+k_3^2) \nonumber \\
& \times (\alpha_B^2+2 \alpha_B-\beta_K+1)^2 (2 f_2 (3 H (\alpha_B+1) (2 \alpha_H-1) (\alpha_B+3 \alpha_H+1)+H ((2 \alpha_H-11) \alpha_B^2\nonumber \\
&+(6 \alpha_H^2-17 \alpha_H-18) \alpha_B+6 \alpha_H^2-19 \alpha_H+\beta_K-7))-f_1 (2 \alpha_B (\alpha_H-1)+2 \alpha_H+\beta_K-2)) a^4\nonumber \\
&-18 f_2 H^2 k_1^2 (k_1^2+k_2^2-k_3^2) (k_1^2-k_2^2+k_3^2) (\alpha_B+1) \alpha_H (\alpha_B^2+2 \alpha_B-\beta_K+1)^2 a^2-18 f_2 H^2 k_2^2 \nonumber \\
& \times (k_1^2+k_2^2-k_3^2) (-k_1^2+k_2^2+k_3^2) (\alpha_B+1) \alpha_H (\alpha_B^2+2 \alpha_B-\beta_K+1)^2 a^2-18 f_2 H^2 k_3^2 (k_1^2-k_2^2\nonumber \\
&+k_3^2) (-k_1^2+k_2^2+k_3^2) (\alpha_B+1) \alpha_H (\alpha_B^2+2 \alpha_B-\beta_K+1)^2 a^2+18 f_2 H^2 k_1^4 (k_1^2-k_2^2-k_3^2) (\alpha_B\nonumber \\
&+1) \alpha_H (6 \alpha_H-1) (\alpha_B^2+2 \alpha_B-\beta_K+1)^2 a^2+9 f_2 H^2 k_3^2 (k_1^2+k_2^2-k_3^2)^2 (\alpha_B+1) (1-3 \alpha_H) \nonumber \\
& \times (\alpha_B^2+2 \alpha_B-\beta_K+1)^2 a^2+9 f_2 H^2 k_2^2 (k_1^2-k_2^2+k_3^2)^2 (\alpha_B+1) (1-3 \alpha_H) (\alpha_B^2+2 \alpha_B-\beta_K\nonumber \\
&+1)^2 a^2+9 f_2 H^2 k_1^2 (-k_1^2+k_2^2+k_3^2)^2 (\alpha_B+1) (1-3 \alpha_H) (\alpha_B^2+2 \alpha_B-\beta_K+1)^2 a^2+18 f_2 H^2 \nonumber \\
& \times k_3^4 (k_1^2+k_2^2-k_3^2) (\alpha_B+1) \alpha_H (1-6 \alpha_H) (\alpha_B^2+2 \alpha_B-\beta_K+1)^2 a^2+18 f_2 H^2 k_2^4 (k_1^2-k_2^2+k_3^2) \nonumber \\
& \times (\alpha_B+1) \alpha_H (1-6 \alpha_H) (\alpha_B^2+2 \alpha_B-\beta_K+1)^2 a^2-18 f_2 H^2 k_1^2 (k_1^2-k_2^2-k_3^2) (k_1^2+k_2^2-k_3^2) \nonumber \\
& \times (\alpha_B+1) (3 \alpha_H^2-\alpha_H+1) (\alpha_B^2+2 \alpha_B-\beta_K+1)^2 a^2-18 f_2 H^2 k_1^2 (k_1^2-k_2^2-k_3^2) (k_1^2-k_2^2\nonumber \\
&+k_3^2) (\alpha_B+1) (3 \alpha_H^2-\alpha_H+1) (\alpha_B^2+2 \alpha_B-\beta_K+1)^2 a^2+18 f_2 H^2 k_2^2 (k_1^2+k_2^2-k_3^2) (k_1^2-k_2^2\nonumber \\
&+k_3^2) (\alpha_B+1) (3 \alpha_H^2-\alpha_H+1) (\alpha_B^2+2 \alpha_B-\beta_K+1)^2 a^2+18 f_2 H^2 k_3^2 (k_1^2+k_2^2-k_3^2) (k_1^2-k_2^2\nonumber \\
&+k_3^2) (\alpha_B+1) (3 \alpha_H^2-\alpha_H+1) (\alpha_B^2+2 \alpha_B-\beta_K+1)^2 a^2-18 f_2 H^2 k_2^2 (-k_1^2+k_2^2-k_3^2) (-k_1^2\nonumber \\
&+k_2^2+k_3^2) (\alpha_B+1) (3 \alpha_H^2-\alpha_H+1) (\alpha_B^2+2 \alpha_B-\beta_K+1)^2 a^2-18 f_2 H^2 k_3^2 (-k_1^2-k_2^2+k_3^2) \nonumber \\
& \times (-k_1^2+k_2^2+k_3^2) (\alpha_B+1) (3 \alpha_H^2-\alpha_H+1) (\alpha_B^2+2 \alpha_B-\beta_K+1)^2 a^2-18 f_2 H^2 k_1^2 k_2^2 (k_1^2-k_2^2\nonumber \\
&-k_3^2) (\alpha_B+1)^2 (\alpha_B+3 \alpha_H+1) (2 \alpha_H^2+\alpha_H-1) (\alpha_B^2+2 \alpha_B-\beta_K+1) a^2-18 f_2 H^2 k_1^2 k_3^2 (k_1^2\nonumber \\
&-k_2^2-k_3^2) (\alpha_B+1)^2 (\alpha_B+3 \alpha_H+1) (2 \alpha_H^2+\alpha_H-1) (\alpha_B^2+2 \alpha_B-\beta_K+1) a^2-18 f_2 H^2 k_2^2 k_3^2 \nonumber \\
& \times (-k_1^2+k_2^2-k_3^2) (\alpha_B+1)^2 (\alpha_B+3 \alpha_H+1) (2 \alpha_H^2+\alpha_H-1) (\alpha_B^2+2 \alpha_B-\beta_K+1) a^2 \nonumber \\
&+18 f_2 H^2 k_1^2 k_3^2 (k_1^2+k_2^2-k_3^2) (\alpha_B+1)^2 (\alpha_B+3 \alpha_H+1) (2 \alpha_H^2+\alpha_H-1) (\alpha_B^2+2 \alpha_B-\beta_K+1) a^2\nonumber \\
&+18 f_2 H^2 k_2^2 k_3^2 (k_1^2+k_2^2-k_3^2) (\alpha_B+1)^2 (\alpha_B+3 \alpha_H+1) (2 \alpha_H^2+\alpha_H-1) (\alpha_B^2+2 \alpha_B-\beta_K+1) a^2\nonumber \\
&+18 f_2 H^2 k_1^2 k_2^2 (k_1^2-k_2^2+k_3^2) (\alpha_B+1)^2 (\alpha_B+3 \alpha_H+1) (2 \alpha_H^2+\alpha_H-1) (\alpha_B^2+2 \alpha_B-\beta_K+1) a^2\nonumber \\
&-6 H k_1^2 k_2^2 (k_1^2-k_2^2-k_3^2) (\alpha_B+1) (\alpha_B^2+2 \alpha_B-\beta_K+1) (3 f_1 (\alpha_B+1) (\alpha_H+1)+f_2 H (18 \alpha_H^3\nonumber \\
&+79 \alpha_H^2-3 \beta_K \alpha_H-7 \alpha_H+2 \alpha_B^2 (7 \alpha_H-4)+\alpha_B (18 \alpha_H^3+79 \alpha_H^2+7 \alpha_H+6)-6 \beta_K+14)) a^2\nonumber \\
&-6 H k_1^2 k_3^2 (k_1^2-k_2^2-k_3^2) (\alpha_B+1) (\alpha_B^2+2 \alpha_B-\beta_K+1) (3 f_1 (\alpha_B+1) (\alpha_H+1)+f_2 H (18 \alpha_H^3\nonumber \\
&+79 \alpha_H^2-3 \beta_K \alpha_H-7 \alpha_H+2 \alpha_B^2 (7 \alpha_H-4)+\alpha_B (18 \alpha_H^3+79 \alpha_H^2+7 \alpha_H+6)-6 \beta_K+14)) a^2\nonumber \\
&+6 H k_1^2 k_3^2 (k_1^2+k_2^2-k_3^2) (\alpha_B+1) (\alpha_B^2+2 \alpha_B-\beta_K+1) (3 f_1 (\alpha_B+1) (\alpha_H+1)+f_2 H (18 \alpha_H^3\nonumber \\
&+79 \alpha_H^2-3 \beta_K \alpha_H-7 \alpha_H+2 \alpha_B^2 (7 \alpha_H-4)+\alpha_B (18 \alpha_H^3+79 \alpha_H^2+7 \alpha_H+6)-6 \beta_K+14)) a^2\nonumber \\
&+6 H k_2^2 k_3^2 (k_1^2+k_2^2-k_3^2) (\alpha_B+1) (\alpha_B^2+2 \alpha_B-\beta_K+1) (3 f_1 (\alpha_B+1) (\alpha_H+1)+f_2 H (18 \alpha_H^3\nonumber \\
&+79 \alpha_H^2-3 \beta_K \alpha_H-7 \alpha_H+2 \alpha_B^2 (7 \alpha_H-4)+\alpha_B (18 \alpha_H^3+79 \alpha_H^2+7 \alpha_H+6)-6 \beta_K+14)) a^2\nonumber \\
&+6 H k_1^2 k_2^2 (k_1^2-k_2^2+k_3^2) (\alpha_B+1) (\alpha_B^2+2 \alpha_B-\beta_K+1) (3 f_1 (\alpha_B+1) (\alpha_H+1)+f_2 H (18 \alpha_H^3\nonumber \\
&+79 \alpha_H^2-3 \beta_K \alpha_H-7 \alpha_H+2 \alpha_B^2 (7 \alpha_H-4)+\alpha_B (18 \alpha_H^3+79 \alpha_H^2+7 \alpha_H+6)-6 \beta_K+14)) a^2\nonumber \\
&+6 H k_2^2 k_3^2 (k_1^2-k_2^2+k_3^2) (\alpha_B+1) (\alpha_B^2+2 \alpha_B-\beta_K+1) (3 f_1 (\alpha_B+1) (\alpha_H+1)+f_2 H (18 \alpha_H^3\nonumber \\
&+79 \alpha_H^2-3 \beta_K \alpha_H-7 \alpha_H+2 \alpha_B^2 (7 \alpha_H-4)+\alpha_B (18 \alpha_H^3+79 \alpha_H^2+7 \alpha_H+6)-6 \beta_K+14)) a^2\nonumber \\
&-2 k_1^2 k_2^2 k_3^2 (\alpha_B+1) ((-96 f_{2,{\rm xxx}} H^2+3 (3 f_1 (\alpha_H+1) (2 \alpha_H^2-7 \alpha_H+\alpha_B (\alpha_H-3)-13)\nonumber \\
&-18 f_2 H (\alpha_B+3 \alpha_H+1) (5 \alpha_H^2+10 \alpha_H-4 \beta_H-3)+6 f_2 H (47 \alpha_H^3+142 \alpha_H^2+(6 \beta_B-48 \beta_H\nonumber \\
&-6 \beta_K+29) \alpha_H+10 \alpha_B^2 (\alpha_H+1)+2 \alpha_B (29 \alpha_H^2+58 \alpha_H+\beta_B-20 \beta_H+5)-6 (6 \beta_H+\beta_K\nonumber \\
&+7))-8 f_{1,{\rm xxx}}) H-8 f_{0,{\rm xxx}}) (\alpha_B+1)^2+9 H (f_1 (-\alpha_H^2+2 \alpha_H+3)+2 f_2 H (35 \alpha_H^2+8 \alpha_H\nonumber \\
&+8 \alpha_B (\alpha_H-2)-2 \beta_B+4 \beta_H-19)) (\alpha_B^2+2 \alpha_B-\beta_K+1) (\alpha_B+1)-54 f_2 H^2 (3 \alpha_H-1) \nonumber \\
& \times ((\alpha_B+1)^2-\beta_K)^2) a^2-4 f_2 k_1^2 k_2^2 k_3^4 (\alpha_B+1)^3 \alpha_H (9 \alpha_H^2-14 \alpha_H+12 \beta_H+1)\nonumber \\
&-4 f_2 k_1^2 k_2^4 k_3^2 (\alpha_B+1)^3 \alpha_H (9 \alpha_H^2-14 \alpha_H+12 \beta_H+1)-4 f_2 k_1^4 k_2^2 k_3^2 (\alpha_B+1)^3 \alpha_H (9 \alpha_H^2-14 \alpha_H\nonumber \\
&+12 \beta_H+1)-4 f_2 k_1^2 k_2^2 k_3^2 (k_1^2-k_2^2-k_3^2) (\alpha_B+1)^3 (3 \alpha_H^3-17 \alpha_H^2+12 \beta_H \alpha_H+\alpha_H-3)\nonumber \\
&+4 f_2 k_1^2 k_2^2 k_3^2 (k_1^2+k_2^2-k_3^2) (\alpha_B+1)^3 (3 \alpha_H^3-17 \alpha_H^2+12 \beta_H \alpha_H+\alpha_H-3)\nonumber \\
&+4 f_2 k_1^2 k_2^2 k_3^2 (k_1^2-k_2^2+k_3^2) (\alpha_B+1)^3 (3 \alpha_H^3-17 \alpha_H^2+12 \beta_H \alpha_H+\alpha_H-3)\Big]
\end{align}

\section{$K^2$ functions}
\label{app2}
In this appendix we give the coefficients of the differential operator leading to the mode functions. 
\begin{align}
K_{\lambda}^2&(x,k)= \frac{\left(c H+\log \left(-\frac{H x}{k}\right)\right)^2}{864 (\alpha_B+1)^2 f_2 H^6 x^4 \left(\alpha_B^2+2 \alpha_B-\beta_K+1\right)^2}    \nonumber \\
& \times \large[ c^2 H^2 (135 (\alpha_B^2+2 \alpha_B-\beta_K+1)^3+(\alpha_B+1)^2 x^4 (4 \alpha_B^2 (3 \alpha_H+1)+\alpha_B (3 \alpha_H^3-3 \alpha_H^2  \nonumber \\
&+17 \alpha_H+7)-3 \alpha_H^2 (\beta_K+1)+3 \alpha_H^3+\alpha_H (5-6 \beta_K)-3 \beta_K+3)+9 (\alpha_B+1) x^2 (\alpha_B^2 \nonumber \\
&+2 \alpha_B-\beta_K+1) (\alpha_B^2 (\alpha_H-5)+\alpha_B (2 \alpha_H+\beta_K-10)+2 \alpha_H \beta_K+\alpha_H+3 \beta_K-5))  \nonumber \\
&+2 \log (-\frac{H x}{k}) (c H (135 (\alpha_B^2+2 \alpha_B-\beta_K+1)^3+(\alpha_B+1)^2 x^4 (4 \alpha_B^2 (3 \alpha_H+1) \nonumber \\
&+\alpha_B (3 \alpha_H^3-3 \alpha_H^2+17 \alpha_H+7)-3 \alpha_H^2 (\beta_K+1)+3 \alpha_H^3+\alpha_H (5-6 \beta_K)-3 \beta_K+3)  \nonumber \\
&+9 (\alpha_B+1) x^2 (\alpha_B^2+2 \alpha_B-\beta_K+1) (\alpha_B^2 (\alpha_H-5)+\alpha_B (2 \alpha_H+\beta_K-10)+2 \alpha_H \beta_K  \nonumber \\
&+\alpha_H+3 \beta_K-5))-9 (\alpha_B^2+2 \alpha_B-\beta_K+1) (21 (\alpha_B^2+2 \alpha_B-\beta_K+1)^2-(\alpha_B+1) x^2  \nonumber \\
&\times (4 \alpha_B^2+\alpha_B (-3 \alpha_H^2+6 \alpha_H+9)-3 \alpha_H^2-2 \alpha_H (\beta_K-3)-2 \beta_K+5)))-18 c H (\alpha_B^2  \nonumber \\
&+2 \alpha_B-\beta_K+1) (21 (\alpha_B^2+2 \alpha_B-\beta_K+1)^2-(\alpha_B+1) x^2 (4 \alpha_B^2+\alpha_B (-3 \alpha_H^2+6 \alpha_H+9)  \nonumber \\
&-3 \alpha_H^2-2 \alpha_H (\beta_K-3)-2 \beta_K+5))+\log ^2(-\frac{H x}{k}) (135 (\alpha_B^2+2 \alpha_B-\beta_K+1)^3  \nonumber \\
&+(\alpha_B+1)^2 x^4 (4 \alpha_B^2 (3 \alpha_H+1)+\alpha_B (3 \alpha_H^3-3 \alpha_H^2+17 \alpha_H+7)-3 \alpha_H^2 (\beta_K+1)+3 \alpha_H^3  \nonumber \\
&+\alpha_H (5-6 \beta_K)-3 \beta_K+3)+9 (\alpha_B+1) x^2 (\alpha_B^2+2 \alpha_B-\beta_K+1) (\alpha_B^2 (\alpha_H-5) \nonumber \\
&+\alpha_B (2 \alpha_H+\beta_K-10)+2 \alpha_H \beta_K+\alpha_H+3 \beta_K-5))+18 (\alpha_B^2+2 \alpha_B-\beta_K+1) (9 (\alpha_B^2  \nonumber \\
&+2 \alpha_B-\beta_K+1)^2+(\alpha_B+1)^2 (3 \alpha_H^2-6 \alpha_H-1) x^2)  \large ] \,.
\end{align}

\begin{align}
K_{m^2}^2&(x,k)=\frac{1}{72 (\alpha_B+1)^2 f_2 H^4 x^4 \left(\alpha_B^2+2 \alpha_B-\beta_K+1\right)^2}   \nonumber \\
&\times \large[ c^2 H^2 (135 (\alpha_B^2+2 \alpha_B-\beta_K+1)^3+(\alpha_B+1)^2 x^4 (4 \alpha_B^2 (3 \alpha_H+1)+\alpha_B (3 \alpha_H^3 -3 \alpha_H^2 \nonumber \\
&+17 \alpha_H+7)-3 \alpha_H^2 (\beta_K+1)+3 \alpha_H^3+\alpha_H (5-6 \beta_K)-3 \beta_K+3)+9 (\alpha_B+1) x^2 (\alpha_B^2 \nonumber \\
&+2 \alpha_B-\beta_K+1) (\alpha_B^2 (\alpha_H-5)+\alpha_B (2 \alpha_H+\beta_K-10)+2 \alpha_H \beta_K+\alpha_H+3 \beta_K-5)) \nonumber \\ 
&+\log (-\frac{H x}{k}) (2 c H (135 (\alpha_B^2+2 \alpha_B-\beta_K+1)^3+(\alpha_B+1)^2 x^4 (4 \alpha_B^2 (3 \alpha_H+1) \nonumber \\
&+\alpha_B (3 \alpha_H^3-3 \alpha_H^2+17 \alpha_H+7)-3 \alpha_H^2 (\beta_K+1)+3 \alpha_H^3+\alpha_H (5-6 \beta_K)-3 \beta_K+3) \nonumber \\
&+9 (\alpha_B+1) x^2 (\alpha_B^2+2 \alpha_B-\beta_K+1) (\alpha_B^2 (\alpha_H-5)+\alpha_B (2 \alpha_H+\beta_K-10)+2 \alpha_H \beta_K \nonumber \\
&+\alpha_H+3 \beta_K-5))-9 (\alpha_B^2+2 \alpha_B-\beta_K+1) (21 (\alpha_B^2+2 \alpha_B-\beta_K+1)^2\nonumber \\
&-(\alpha_B+1) x^2 (4 \alpha_B^2+\alpha_B (-3 \alpha_H^2+6 \alpha_H+9)-3 \alpha_H^2-2 \alpha_H (\beta_K-3)-2 \beta_K+5)))\nonumber \\
&-9 c H (\alpha_B^2+2 \alpha_B-\beta_K+1) (21 (\alpha_B^2+2 \alpha_B-\beta_K+1)^2-(\alpha_B+1) x^2 (4 \alpha_B^2+\alpha_B (-3 \alpha_H^2 \nonumber \\
&+6 \alpha_H+9)-3 \alpha_H^2-2 \alpha_H (\beta_K-3)-2 \beta_K+5))+\log ^2(-\frac{H x}{k}) (135 (\alpha_B^2+2 \alpha_B-\beta_K+1)^3 \nonumber \\
&+(\alpha_B+1)^2 x^4 (4 \alpha_B^2 (3 \alpha_H+1)+\alpha_B (3 \alpha_H^3-3 \alpha_H^2+17 \alpha_H+7)-3 \alpha_H^2 (\beta_K+1) \nonumber \\
&+3 \alpha_H^3+\alpha_H (5-6 \beta_K)-3 \beta_K+3)+9 (\alpha_B+1) x^2 (\alpha_B^2+2 \alpha_B-\beta_K+1) (\alpha_B^2 (\alpha_H-5)\nonumber \\
&+\alpha_B (2 \alpha_H+\beta_K-10)+2 \alpha_H \beta_K+\alpha_H+3 \beta_K-5))+3 (\alpha_B^2+2 \alpha_B-\beta_K+1) (9 (\alpha_B^2 \nonumber \\
&+2 \alpha_B-\beta_K+1)^2+(\alpha_B+1)^2 (3 \alpha_H^2-6 \alpha_H-1) x^2)   \large] \,.
\end{align}

\acknowledgments
We acknowledge the use of the \textsc{xPand} package \footnote{http://www.xact.es/xPand/} \cite{Pitrou:2013hga} for computing the perturbations. 
AL acknowledges funding by the LabEx ENS-ICFP: ANR-10-LABX-0010/ANR-10-IDEX-0001-02 PSL*. This project has received funding /support from the European Union’s Horizon 2020 research and innovation programme under the Marie Skłodowska-Curie grant agreement No 860881-HIDDeN.

\bibliography{Bibliography}{}

\end{document}